\documentclass[journal ]{new-aiaa}

\usepackage{placeins}
\usepackage[table]{xcolor}
\usepackage{multirow}
\usepackage{graphicx}
\usepackage[utf8]{inputenc}
\usepackage{textcomp}
\usepackage{graphicx}
\usepackage{amsmath}
\usepackage{siunitx}
\usepackage{longtable,tabularx}
\usepackage{subfig}
\usepackage{subfloat}

\newcommand{\bbm}{\begin{bmatrix}}
\newcommand{\ebm}{\end{bmatrix}}
\newcommand{\vect}{\underrightarrow}
\DeclareRobustCommand{\vect}{\underrightarrow}
\DeclareMathAlphabet{\mbf}{OT1}{ptm}{b}{n}
\newcommand{\mbs}[1]{{\boldsymbol{#1}}}
\newcommand{\mbshat}[1]{{\hat{\boldsymbol{#1}}}}
\newcommand{\norm}[1]{\left\lVert#1\right\rVert} 
\newcommand{\trans}{{\ensuremath{\mathsf{T}}}} 
\newcommand{\p}{\partial}

\newcommand{\onehalf}{\mbox{$\textstyle{\frac{1}{2}}$}}

\newcommand{\beq}{\begin{equation}}
\newcommand{\eeq}{\end{equation}}
\newcommand{\bdis}{\begin{displaymath}}
\newcommand{\edis}{\end{displaymath}}

\title{Static and Dynamic Torque Generation Analysis of a Cable-Actuated Solar Sail}
\author{Keegan R. Bunker\footnote{Ph.D. Student, Department of Aerospace Engineering \& Mechanics, 110 Union St. SE, Minneapolis, MN 55455, email: bunke029@umn.edu} and Ryan J. Caverly\footnote{Associate Professor, Department of Aerospace Engineering \& Mechanics, 110 Union St. SE, Minneapolis, MN 55455, AIAA Member, email: rcaverly@umn.edu (Corresponding Author)}}
\affil{University of Minnesota, Twin Cities, Minneapolis, MN 55455}
\begin{document}

\maketitle

\section{Introduction}
\lettrine{S}{olar} sails are large space structures consisting of a lightweight sail that utilizes solar radiation pressure (SRP) to produce usable thrust for a spacecraft without requiring fuel of any kind. Though small, this thrust can be used over long durations to enable unique spacecraft mission designs, including interstellar missions~\cite{gong2019review} and orbits out of the ecliptic plane~\cite{thomas2020solar,kobayashi2020high}. If chemical fuel is not used for any other systems onboard the spacecraft, a solar sail can increase the spacecraft operational life span immensely. LightSail~2 showed the potential of solar sails \cite{spencer2021lightsail}, raising its orbital altitude in low Earth orbit from the thrust generated by the sail over the course of its mission. The space community continues to see value in developing solar sails further, as seen by the launch of the Advanced Composite Solar Sail System (ACS3) technology exploration mission into low Earth orbit in to test novel solar sail boom technologies~\cite{wilkie2021overview}.

\par Attitude control of solar sails is key to meeting mission requirements, motivating most previous and proposed missions to use reaction wheels as actuators. A typical solar sail attitude control system with reaction wheels includes spinning metal wheels onboard the spacecraft that can generate torques along any direction through the conservation of angular momentum by altering their spin rate. The spin rate has a limited operating range, so another actuator is required to apply a net torque to the spacecraft and allow the reaction wheel speeds to be adjusted without affecting the attitude. This capability is called momentum management. The momentum management requirements will become more challenging as solar sails of substantially larger size are considered for future missions~\cite{thomas2020solar,Berthet2024-mi}, thus requiring larger reaction wheels. Moreover, larger solar sails will likely have more structural flexibility~\cite{pimienta2019heliogyro,brownell2023time,Boni2023-mb}, resulting in unwanted structural deformations and significant disturbance torques due to imbalances in SRP~\cite{gauvain2023solar}.

\par Spacecraft missions outside of low Earth orbit typically rely on attitude control system (ASC) thrusters for momentum management, which has also been proposed for solar sails~\cite{inness2023momentum,wie2007solar}. ACS thrusters use chemical propulsion to apply thrust and generate net torques on the spacecraft. ACS thrusters negate the fuel-free benefits of a solar sail and limit the operational life of a solar sail spacecraft. Several alternative technologies have been investigated to address solar sail momentum management demands without fuel. An effective means to generate momentum management torques that has been extensively explored involves shifting the solar sail's center of mass using an onboard translation mechanism~\cite{wie2007solar,bolle2008solar,lappas2009practical,romagnoli2011high,adeli2011scalable,huang2019solar,firuzi2018attitude, Chujo2024-nm}. NASA's Solar Cruiser is equipped with a version of this known as an active mass translator (AMT), which allows for the relative displacement of the upper and lower halves of the solar sail bus~\cite{inness2023momentum}. The AMT is capable of generating torques by displacing the center of mass relative to the center of SRP in two axes for momentum management, although it is unclear how this technology will scale up to solar sails with much larger areas due to limits in its range of actuation. Gimbaled ballast masses have also been proposed as a means to shift the solar sail's center of mass using a similar concept~\cite{wie2004solar2,sperber2016large}, however, they have remained largely theoretical and may increase the overall mass of the solar sail if the added mass is truly ballast. 

A variety of concepts involving adjusting the reflectivity properties of portions of the solar sail in order to generate a torque through an imbalance in SRP~\cite{funase2012modeling,borggrafe2014optical,ullery2018strong,davoyan2021photonic,Qi2024-ls} have been proposed. This concept is being considered for use on Solar Cruiser in the form of reflectivity control devices (RCDs) angled from the normal of the solar sail membrane to generate torques out of the plane of the sail~\cite{inness2023momentum}. Although this is a promising technology, it produces relatively small torques and comes with the challenge of providing power to devices far from the solar sail bus. 

Control vanes at the tips of the solar sail's structural booms have been proposed as a means to generate torques similar to the control surfaces on an aircraft~\cite{choi2015structural,eldad2017minimum,hassanpour2020collocated,eldad2015propellantless,choi2016control}. Unfortunately, a significant challenge in the use of control vanes is actuating them (most likely with some sort of motor) at the end of long, flexible booms. The control vanes also need to have significant area in order to generate significant torques. 

In an effort to generate large torques with minimal actuation, it has been proposed to angle the panels of a solar sail~\cite{Chujo2022-oq,guerrant2015tactics}. Although this can create large changes in torques, it is only effective when fairly rigid solar sail panels/membranes are used and does not scale well to large lightweight sails. A more practical adaptation of this for large solar sails is the idea of actively shifting the attachment point of the sail membrane quadrant at the end of the boom tip~\cite{fu2015attitude}. This can provide large changes in torque, although this also features the challenge of actuation far from the spacecraft bus and can significantly reduce the SRP thrust during actuation. A more detailed description of many of these actuators and other concepts can be found in~\cite{gong2019review,fu2016solar}.

\par
\begin{figure}[t!]
    \centering
    \includegraphics[width=0.48\textwidth]{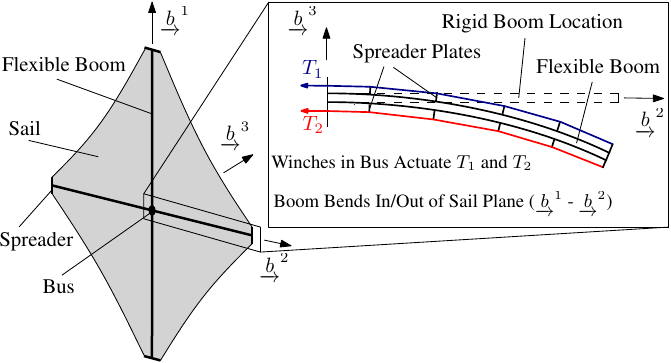} 
    \caption{An overview of the CABLESSail concept. Tensioning cables running through spreader plates along the booms deform the booms and sail shape.}
    \label{fig:CABLESSailConcept}
\end{figure}

\begin{figure}[t!]
    \centering
    \includegraphics[width=0.48\textwidth]{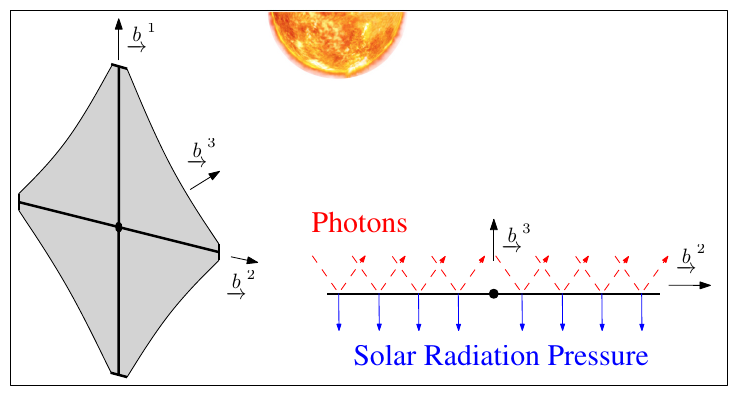}  
    \includegraphics[width=0.48\textwidth]{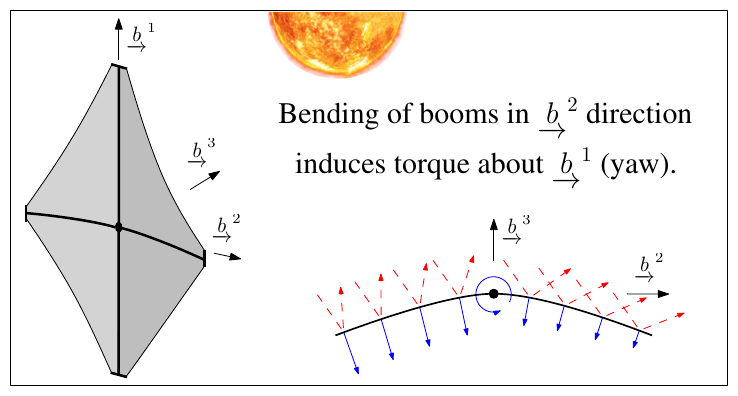} 
    \caption{Torque generation about the pitch/yaw axes of CABLESSail by bending of the booms. The photon incidence angle on each half of the sail is changed and an imbalance in SRP generates the yaw torque.}
    \label{fig:TorqueDemo}
\end{figure}

 The Cable-Actuated Bio-inspired Elastic Solar Sail (CABLESSail) concept, shown in Fig.~\ref{fig:CABLESSailConcept},  utilizes controlled deformation of the solar sail's flexible boom structure to change the shape of the sail membrane and produce usable momentum management torques from a differential in SRP. In a sense, CABLESSail aims to transform the undesirable flexible deformations of large solar sails into controlled desirable deformations to generate usable momentum management torques. Several cables run from the spacecraft bus along each boom, through spreader plates affixed to the booms at set increments, and are fastened to a plate at the end of each boom. Winches within the spacecraft bus pull on these cables to apply tension, which deforms the booms. With coordinated bending, an SRP differential on the sail can generate usable momentum management torques, which is demonstrated in Fig.~\ref{fig:TorqueDemo}. The torques generated inherently scale with the size of the solar sail, since the deflection of the sail is used to create these torques, without the need for chemical fuel. Further details on the preliminary concept behind the CABLESSail technology can be found in~\cite{caverly2023solar}. This technology shares similarities with the concept in~\cite{zhang2021solar,zhang2021three} that involves the use of piezoelectric actuation to purposefully deform the solar sail's booms. A major distinction between the CABLESSail concept and the concept in~\cite{zhang2021solar,zhang2021three} is that cable actuation has the ability to generate much more substantial boom deformations. Additionally, CABLESSail's actuating cables can hold the boom in a deformed shape without the use of power through a motor brake, unlike the constant voltage needed with piezoelectric actuation.

\par Three key performance areas are investigated in this note to determine the preliminary capabilities of CABLESSail. Firstly, control of flexible space structures is a challenging area of research~\cite{papadopoulos2021robotic} and cable-actuated control is even more of a nascent field, so a preliminary analysis of controlled boom tip deflection via basic open-loop cable tensioning is undertaken to assess the potential performance of future feedback boom tip control. The capability of CABLESSail to generate body torques must be comparable to existing technologies to be a worthwhile alternative; therefore, for the second analysis, CABLESSail is directly compared to an AMT. Uncertainty in the shape of the sail membrane has been shown to significantly impact the SRP disturbance torques acting on solar sails in the presence of undesirable boom deflections and is a driver for attitude and momentum control requirements~\cite{gauvain2023solar}. CABLESSail purposefully makes use of boom deflections to generate SRP torques, so the ability to predictably alter the SRP torque under a wide range of sail membrane shape conditions with coordinated boom bending is examined under the final analysis. 
The first two analyses consist of dynamic simulations of CABLESSail and the AMT example solar sail, while the third is a static analysis of CABLESSail assuming steady state control of the boom tips has been achieved.

\par The novel contribution of this work is a preliminary demonstration of CABLESSail's capabilities through controlled boom tip deflection. Specifically, these include CABLESSail's ability to 1) generate a larger attitude change in comparison to an AMT, 2) reliably generate large pitch and yaw torques  under solar sail membrane shape uncertainty, and 3) reliably generate a large roll torque with some residual pitch and yaw torques under solar sail membrane shape uncertainty. The remainder of this note proceeds with a description of important notation, the definition of Sun incidence angle (SIA) and clock angle, and a summary of the dynamic modeling approach. Four numerical simulation test cases with accompanying results are given in Section~\ref{sec:Numerical simulation}. Concluding remarks follow in Section~\ref{sec:Conclusion}.

\section{Preliminaries} \label{sec:Prelims}

\par This section provides information regarding notation used throughout the note, definitions of the SIA and the clock angle, an overview of the dynamic model of the solar sail structure, and the SRP solar sail membrane modeling.

\subsection{Notation} \label{sec:Notation}
  
\par The following notation is used throughout this note. A reference frame of three orthonormal basis vectors, $\vect{a}^1, \vect{a}^2, \vect{a}^3$, makes up reference frame $\mathcal{F}_a$. The attitude of reference frame $\mathcal{F}_b$ relative to reference frame $\mathcal{F}_a$ is described through the direction cosine matrix (DCM) $\mathbf{C}_{ba} \in \mathbb{R}^{3 \times 3}$. The DCM is further transcribed into a 3-2-1 Euler angle rotation sequence of principle rotations $\mbf{C}_{ba} = \mbf{C}_1(\phi) \mbf{C}_2(\theta) \mbf{C}_3(\psi)$ where $\phi$, $\theta$, and $\psi$ are yaw, pitch, and roll respectively, to match the convention used on NASA solar sail missions. SRP torques are typically resolved in the solar sail body frame, $\mathcal{F}_b$, where the directions are referred to by their associated Euler angles; yaw, $\vect{b}^1$, pitch, $\vect{b}^2$, and roll, $\vect{b}^3$. Further details regarding the notation used for physical vectors, reference frames, and DCMs are found in~\cite{Hughes2004,DeRuiter2013}.

\subsection{Sun Incidence Angle (SIA) \& Clock Angle} \label{sec:Sun Incidence Angle and Clock Angle}
\par The position of the sun relative to the solar sail membrane is a primary driver of SRP force and SRP torque \cite{gauvain2023solar}. For the work presented here, the direction of the sun pointing vector, $\vect{s}$, that points from the solar sail spacecraft center of mass to the sun is defined by the SIA and the clock angle. SIA is the angle between the sun pointing vector and $\vect{b}^3$, the vector normal to the undeformed solar sail. Clock angle is the angle from $\vect{b}^1$ to the sun pointing vector projected into the solar sail plane defined by $\vect{b}^1$ and $\vect{b}^2$. SIA is always a positive value and the sign of the clock angle follows the right handed-convention about$\vect{b}^3$. The clock angle, SIA, and the body frame vectors, $\vect{b}^1$, $\vect{b}^2$, and $\vect{b}^3$, are shown in Fig.~\ref{fig:assembly and angles}.

\section{Solar Sail Modeling} 
\par Modeling of the solar sail is divided into two parts. The dynamic modeling is summarized with a description of the two solar sail models considered in this work, and the SRP and solar sail membrane shape modeling approach are shown. 

\subsection{Solar Sail Assemblies}
\par
\begin{figure}[t!]
\centering
\subfloat[]{
    \includegraphics[width=0.35\textwidth]{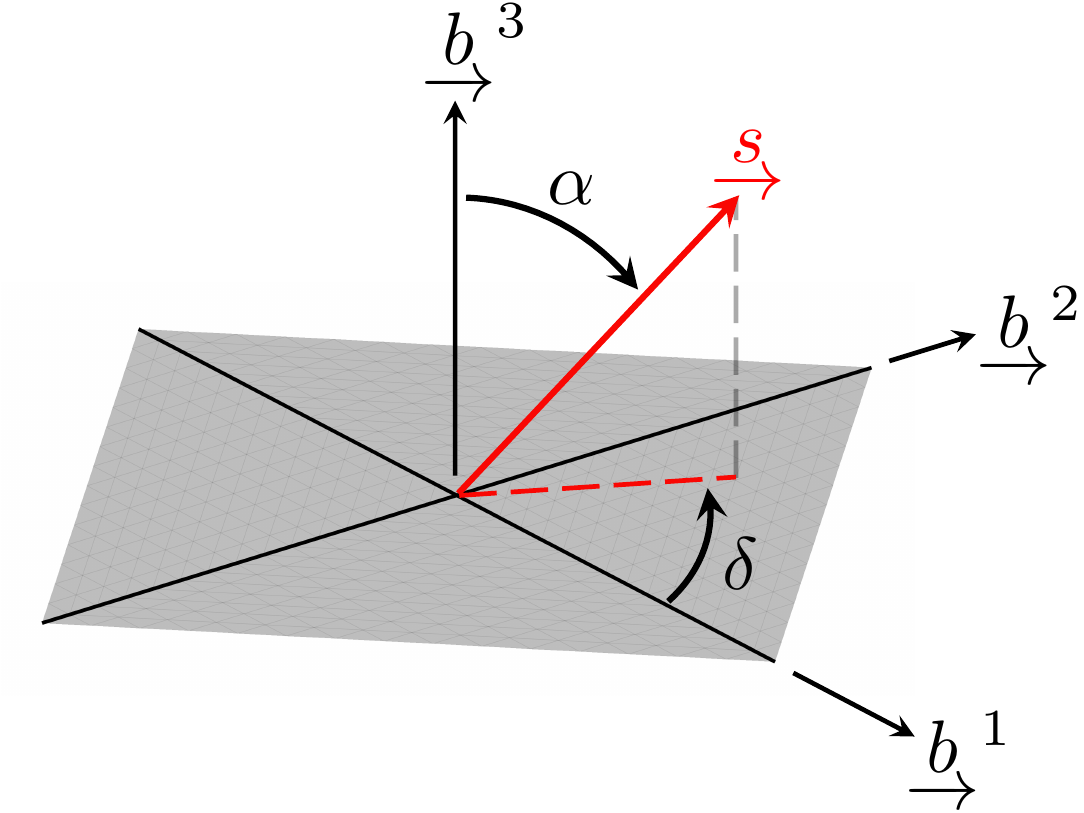}
    \label{fig:second}
}
\hspace{1cm}
\subfloat[]{
    \includegraphics[width=0.5\textwidth]{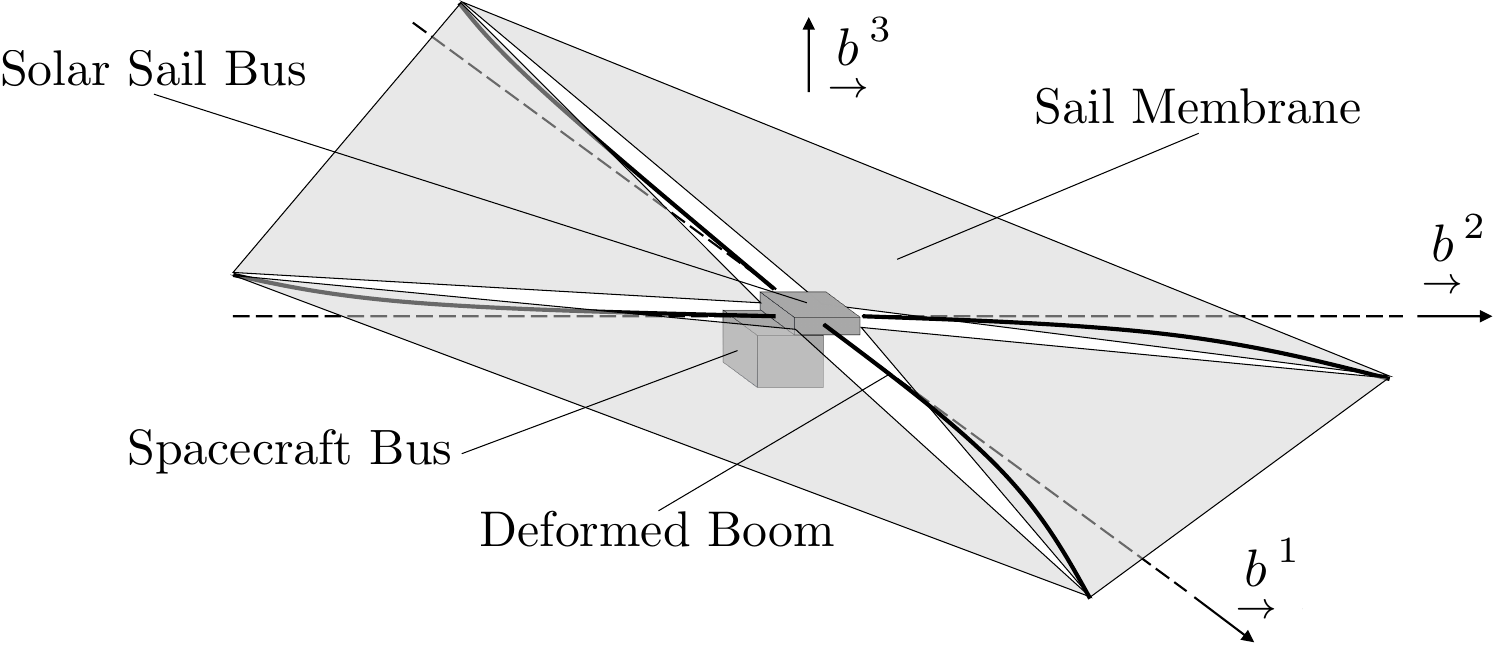}
    \label{fig:DeformedSailAssemblyBothBusses}
}
\caption{(a) Sun incidence angle, $\alpha$, and clock angle, $\delta$, conventions shown with flat sail membranes and straight booms. (b) Solar sail Assembly~\#2 with flat sail membranes, deformed booms, and body frame $\mathcal{F}_b$ attached.}
\label{fig:assembly and angles}
\end{figure}

 Two solar sail models are assembled with this modeling approach for comparison in the work presented here, referred to as Assembly~\#1 and Assembly~\#2. The former is the default CABLESSail design consisting of four boom components and the solar sail bus component. Assembly~\#2 ignores the cable actuation within the booms but incorporates an AMT through a static offset between the spacecraft bus and solar sail bus components that is applied during the assembly step of the dynamic modeling. A depiction of Assembly ~\#2 is shown in Fig.~\ref{fig:DeformedSailAssemblyBothBusses} with the solar sail body frame directions labeled. With zero static offset between the busses of Assembly~\#2, the depiction in Fig.~\ref{fig:DeformedSailAssemblyBothBusses} is identical to Assembly~\#1. Specific material and design values are provided in Section~\ref{sec:Numerical simulation}.

\subsection{Dynamic Modeling Overview} 

\par The dynamic modeling approach used in this work is covered in detail in~\cite{bunker2024modular} and the Appendix, but summarized here for brevity.  Dynamic modeling of the solar sail is accomplished by first modeling the dynamics of each solar sail component individually. These components are the solar sail bus that would house the sail membrane and booms prior to deployment, the four CABLESSail booms that are the primary application of the CABLESSail technology, and the optional spacecraft bus that allows the solar sail bus to be split into two halves for the use of an AMT. The solar sail membrane is not modeled dynamically as a component, rather the SRP force and torque are computed via a solar sail membrane shape model defined in Section~\ref{sec:SolarSailModeling} and applied to the solar sail bus component. The booms are modeled with Euler-Bernoulli beam theory and the assumed modes method, with the cables imparting vertical loads at the location of the spreader plates as tension is applied. The two buses are modeled as uniform rigid bodies. The dynamic equations of motion of each $i$'th component are derived utilizing Lagrange's equations and the associated kinetic and potential energy of each component. Each set of equations are represented in the form$\mbf M_i \dot{\mbs{\nu}}_i + \mbs \Gamma ^\trans _i \mbf K_i \mbf q_i + \mbf f^{\text{non}}_i = \mbs \Gamma_i ^\trans \mbf f_i + \mbs\Xi_i^\trans \mbs\lambda_i$, where $\mbf q_i$ contains the generalized coordinates, $\mbs{\nu}_i$ contains the generalized velocities related to the generalized coordinates by $\dot{\mbf{q}}_i = \mbs{\Gamma}_i\mbs{\nu}_i$, $\mbf M_i$ is the mass matrix, $\mbf K_i$ is the stiffness matrix, $\mbf f^{\text{non}}$ contains all terms that are not linearly dependent on the coordinates and velocities, $\mbf{f}_i$ contains all generalized external forces acting on the component such as the SRP force and torque on the bus and the vertical cable loads on the booms, and $\mbs{\Xi}_i$ and $\mbs{\lambda}_i$ are the constraint matrix and Lagrange multipliers that enforce the attitude parameterization constraint for the $i$'th component. All chosen components are then constrained together to form the entire solar sail according to the chosen solar sail spacecraft design. The equations of motion of all components are concatenated into one multibody equation of motion of the form $\mbf M \dot{\mbs \nu} + \mbs \Gamma^\trans \mbf K \mbf q + \mbf f ^\text{non} = \mbs \Gamma ^\trans \mbf f +\mbs\Xi^\trans \mbs\lambda$, with the assembly constraints enforced through $\mbs{\Xi}$ and $\mbs{\lambda}$ along with the concatenated component attitude parameterization constraints. There are redundant generalized velocities due to the assembly constraints, so a reduced set of velocities, $\hat{\mbs{\nu}}$, is related to the full set of velocities through the transformation matrix $\mbs{\Upsilon}$ by the relation $\mbs \nu = \mbs \Upsilon \hat{\mbs \nu}$, where $\mbs \Xi \mbs \Upsilon = \mbf 0$. Exploiting this relationship, the transformation matrix $\mbs{\Upsilon}$ is used to remove all Lagrange multipliers from the equations of motion resulting in the final dynamic model of the assembly, $\bar{\mbf M} \dot{\hat{\mbs{\nu}}} + \bar{\mbf K}\mbf q +\bar{\mbf{f}}^\text{non} = \bar{\mbf{f}}$, where $\bar{\mbf{M}} = \mbs{\Upsilon}^\trans\mbf{M} \mbs{\Upsilon}$, $\Bar{\mbf{K}} = \mbs{\Upsilon}^\trans\mbs{\Gamma}^{\trans} \mbf{K}$, $\bar{\mbf{f}}_{non} = \mbs{\Upsilon}^\trans\mbf{f}^{non} + \mbs{\Upsilon}^\trans\mbf{M} \dot{\mbs{\Upsilon}} \hat{\mbs{\nu}}$, and $\bar{\mbf{f}} = \mbs{\Upsilon}^\trans\mbs{\Gamma}^{\trans}\mbf{f}$. The final dynamic model used in this work has a reduced state consisting of the position and velocity of the solar sail bus center of mass, the attitude of the solar sail bus, and the deformation states of each boom. The inputs are the state-dependent SRP force and torque covered in the following sub-section, and the cable tensions. For a thorough derivation of the dynamic model, see~\cite{bunker2024modular} and the Appendix.

\subsection{Solar Sail Membrane Modeling}
\label{sec:SolarSailModeling}

\par High-fidelity physics-based dynamic modeling of a sail membrane that considers stretching and applied loads is an open area of research~\cite{Satou2014,Okuizumi2021}. In this note we do not seek to implement a high-fidelity dynamic membrane model, rather we seek to test across assumed static membrane shapes that conservatively represent the effects of static membrane deflection on the SRP torques experienced by a solar sail. The conservative modeling technique that follows is derived from the attitude control analysis for the Solar Cruiser mission at NASA Marshall Space Flight Center~\cite{gauvain2023solar} and consists of a membrane shape model and an SRP force model that considers the optical properties of the sail material. The shape model is modified to enforce the sail membrane boundary conditions that align with the membrane-boom attachment points used in the CABLESSail design, which differ from Solar Cruiser. The force model is unaltered.

\par
\begin{figure}[ht!]
    \centering
    \includegraphics[width=0.45\textwidth]{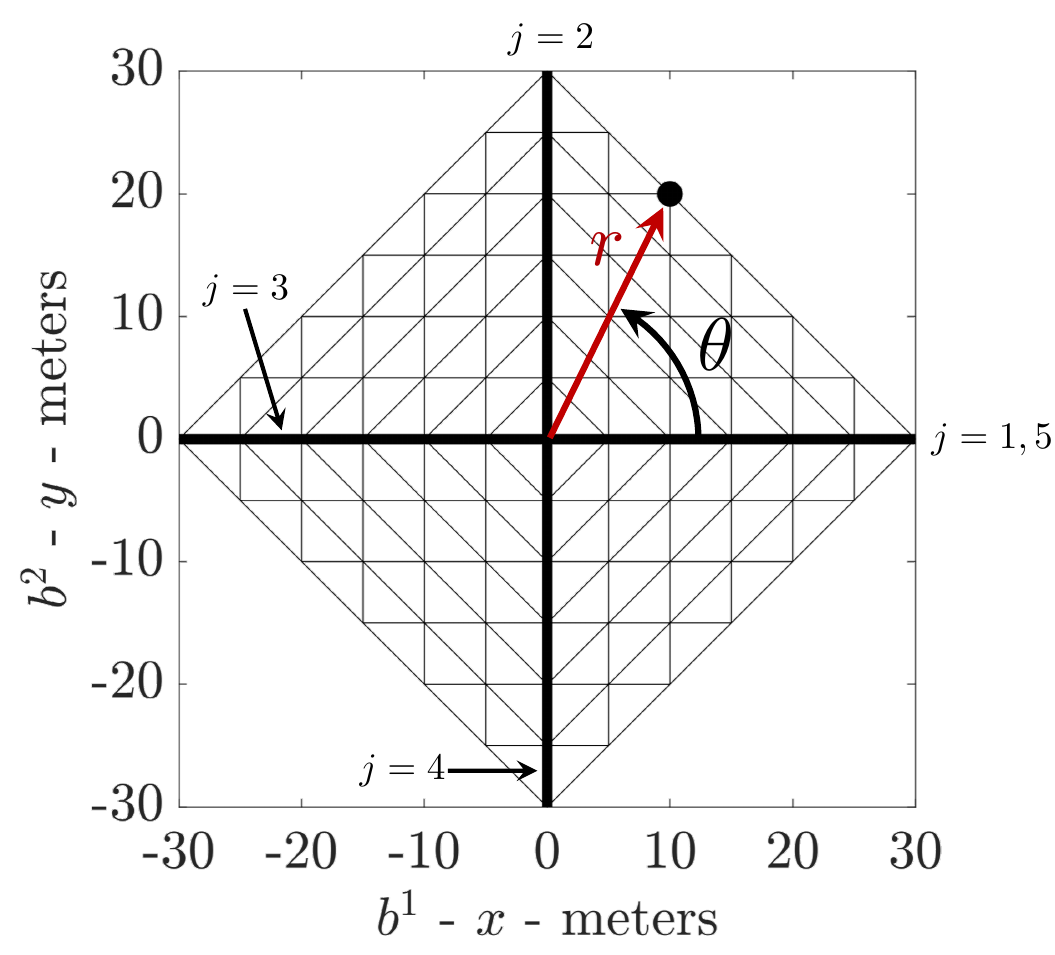 }
    \caption{Top-down view of the triangular solar sail membrane mesh with booms along the $x$ and $y$ directions corresponding to $\vect{b}^1$ and $\vect{b}^2$ directions accordingly.}
    \label{fig: sail membrane coordiantes}
\end{figure}

 The solar sail membrane is modeled as four triangular quadrants, each attached at the two respective boom tips and the solar sail bus center of mass. Nominally the sail is flat with zero deflection out of the plane defined by the three aforementioned boundary constraint points. The continuous surface is represented as a triangular point mesh of a chosen density. The static sail deflection of each mesh point out of this plane is defined by the basis function

\begin{equation} \label{eq: Sail basis function}
    z_{sail} (r,\theta)= \Delta z_{max} 
        \sin \left( \frac{3r \pi \sqrt{2} }{4L }\right)
        \sin \left( \frac{ \theta - \theta_{j} }{ \theta_{j+1} - \theta_{j} }\right)
        ,
\end{equation}
where $r = \sqrt{x^2 + y^2}$,  $\theta = \arctan(y/x)$, L is the boom length, $x$ and $y$ are the locations along the $\vect{b}^1$ and $\vect{b}^2$ directions respectively, $\Delta z_{max}$ is the chosen maximum out of plane sail membrane deflection, and $\theta_j = (j-1)\frac{\pi}{2}$ for $j \in [1,5]$, where $j$ corresponds to the boom number as depicted in Fig. \ref{fig: sail membrane coordiantes}. This numbering convention is also used to reference specific booms for actuation via CABLESSail, although the first boom is double counted to account for the wrapping of the angle $\theta = 0$~rad and $\theta = 2\pi$~rad.  This basis function is a modification from~\cite{gauvain2023solar}. The point of highest deflection due to~\eqref{eq: Sail basis function} occurs at the centroid of the planar sail quadrant similar to observed sail billowing behavior~\cite{gauvain2023solar}. The moments of inertia of the sail membrane about the center of mass of the solar sail bus are added to said bus, along with the total sail mass to approximate the inertia properties. Detailed, qualitative data of in-situ solar sail membrane shape behavior is not readily available. In this work we seek to conduct a preliminary analysis of the capabilities of CABLESSail with an overly-conservative representation of the effects of membrane shape on SRP torque. To that end, maximum membrane deflections up to $ \Delta z_{max = }\pm 15 \text{cm}$ are considered, which is triple of what was considered for the attitude control analysis of Solar Cruiser ($\pm 5 \text{cm}$)~\cite{gauvain2023solar} and double the maximum deflection estimated for the IKAROS solar sail mission~\cite{Satou2014}.Membrane deflection is an uncontrollable effect that induces unintended variations in SRP force and torque, which scale with maximum deflection. The analysis in this note is conservative by analyzing performance under much larger disturbances relative to previous solar sail mission analyses. The area of each sail membrane quadrant is not strictly preserved across membrane and boom tip deflections with this model. The change in membrane area is expected to be minimal and have little impact on the results for the boom lengths and deflection magnitudes considered in this note, especially considering the conservative membrane deflections used in this study.

\par The SRP force model, described in \cite{heaton2015update}, was first proposed at NASA's Jet Propulsion Laboratory in the 1970s and has become a standard solar sail thrust model. The optical coefficients were experimentally re-derived in~\cite{heaton2015update} with modern solar sail membrane materials at NASA Marshall Space Flight Center to support the NASA Near Earth Asteroid Scout and Lunar Flashlight missions. The force exerted on each triangular sail mesh element by the local SRP is applied at the element's planar centroid and modeled as a normal force, $F_n^i$, and tangential force, $F_t^ i$,  given by \cite{heaton2015update}
\begin{align*}
    F_t^i &= PA^i(1 - rs)\cos(\alpha^i)\sin(\alpha^i), 
    \\
    F_n^i &= -PA^i(1 - rs)\cos^2(\alpha^i) -PA^i B_f(1-s)r\cos(\alpha^i) - PA^i(1-r)\cos(\alpha^i) \left(\frac{e_fB_f-e_bB_b}{e_f+e_b} \right), 
\end{align*}
where $P$ is the solar pressure at one astronomical unit (au), $A^i$ is the area of the i'th sail element, $r$ is the reflection coefficient, $s$ is the fraction of specular reflection coefficient, $\alpha^i$ is the local Sun incidence angle (SIA) of the i'th sail element, $B_b$ and $B_f$ are the back and front non-Lambertian coefficients, and $e_b$ and $e_f$ are the back and front surface emissivity, respectively. The sail element area $A^i$ and local Sun incidence angle $\alpha^i$ are dependent on the SIA and clock angle of the whole solar sail, the boom tip deflections, and membrane shape. All other parameters are constant across all sail element and are listed in Table \ref{tab:Solar Sail Optical coefficients} based on the coefficients in~\cite{heaton2015update}. These forces are represented in vector form as $\vect{F}_n^i = F_n^i \vect{n}^i$ and $\vect{F}_t^i = F_t^i \vect{t}^i$, where $\vect n^i$ is the i'th sail element's normal vector, and $\vect{t}^i = -\left(\vect{s} - \vect{s}\cdot \vect{n}^i \vect{n}^i \right)/ \norm{\vect{s} - \vect{s}\cdot \vect{n}^i \vect{n}^i}_2$ is the i'th element's tangential direction which lies in the i'th element plane and is coplanar with $\vect s$, the sun pointing vector. Each $\vect F_n^i$ and $\vect F_t^i$ is applied at their respective element's planar centroid. All forces are resolved in the body frame, $ \mathcal{F}_b$, and represented in the dynamic model by a single force-moment pair applied at the center of mass of the solar sail bus with all respective moment arms due to the respective sail element centroid locations accounted for in the moment calculation. This force and moment are referred to as the SRP force and SRP torque, respectively, representing the total thrust generated by the sail membrane and any torques generated due to misalignment of the SRP center of pressure and spacecraft center of mass. This force and torque are incorporated into the respective generalized force term, $\mbf{f}_i$ and $\bar{\mbf f}$, through their virtual work computations. For more details, see \cite{bunker2024modular}
and the Appendix.

\begin{table}[ht!]
    \caption{Solar sail optical coefficients derived in \cite{heaton2015update} that are used in the numerical simulations of Section~\ref{sec:Numerical simulation}. }
    \label{tab:Solar Sail Optical coefficients}
    \centering
    \begin{tabular}{c c c c c c c c }
        \hline \hline
        \textbf{Coefficient} & P & $r$ & $s$ & $B_b$ & $B_f$ & $e_b$ & $e_f$ \\
        \hline
        \textbf{Value}       & $4.5391 \times 10^{-6} \, \text{N} / \text{m}^2$ & 0.91 & 0.94 & 0.67 & 0.79 & 0.27 & 0.025\\
        \hline \hline
    \end{tabular}   
\end{table}

\section{Numerical Simulation \& Results}
\label{sec:Numerical simulation}

\par Numerical simulations of CABLESSail implementing the modular dynamic modeling approach presented in this note are conducted to investigate the CABLESSail design. First, the impacts of non-flat solar sail membrane shape and boom tip deflections on induced SRP torque is investigated through a static analysis of deflected booms and varying sail membrane shape. 
Next, the ability to control boom tip deflection is tested under a dynamic simulation with mock open-loop tension control. The ability of CABLESSail to alter the spacecraft attitude is compared to existing technology in a dynamic simulation. Finally, the capability of CABLESSail to alter the SRP torque through boom deflection is tested under random sail shape conditions through a Monte Carlo test.

\par Material and design values for the boom components used in Assembly \#1 and Assembly \#2 are listed in Table \ref{tab:Material Constants}. All values are identical to those used in~\cite{bunker2024modular}, except for the flexural rigidity, which is increased for this work to better represent the booms designed for Solar Cruiser~\cite{nguyen2023solar}. Table~\ref{tab:assembly bus properties} contains the dimensions and masses of the buses used for Assembly \#1 and Assembly \#2. Note that the buses for Assembly \#2 account for the same overall mass and dimension as the single bus used for Assembly~\#1, such that both solar sail assemblies have identical inertial properties when the AMT offset between the buses in Assembly~\#2 is zero. The prescribed offset is noted in the relevant test cases.

\begin{table}[ht!]
\caption{Material and component constants used throughout the numerical simulations in Section~\ref{sec:Numerical simulation}. }
\label{tab:Material Constants}
\begin{center}
\begin{tabular}{c l l} 
    \hline \hline
        \textbf{Symbol} & \multicolumn{1}{c}{\textbf{Parameter}} & \multicolumn{1}{c}{\textbf{Value}} \\ [0.5ex] 
    \hline
        $L$ & Boom Length & $29.5$ m \\ 
    \hline
        $R$ & Boom Radius & $0.1$ m  \\
    \hline
        $m_{B_i}$& Boom Mass & 3 kg \\
    \hline
        $\rho$ & Boom Linear Density & $0.1017$ kg/m$^2$   \\ 
    \hline
        $EI$ & Boom Flexural Rigidity & 1700 Nm$^2$ \\
    \hline
        $h$ & Spreader Plate Distance Beyond Boom Radius & $0.1$ m  \\
    \hline
        - & Number of Spreader Plates & 19 \\
    \hline
        $\Delta x$ & Spreader Plate Spacing & 1.5526 m \\
    \hline
        $m_{sail}$ & Solar Sail Membrane Mass & $50$ kg \\
    \hline \hline
\end{tabular}
\end{center}
\end{table}

\begin{table}[ht!]
    \caption{Solar sail bus and spacecraft bus dimensions and mass for Assembly~\#1 and Assembly~\#2.}
    \label{tab:assembly bus properties}
    \centering
    \begin{tabular}{ c c c } 
        \hline \hline
        \multicolumn{1}{c}{}                   
            & \textbf{Solar Sail Bus}& \textbf{Spacecraft Bus} \\
        \hline
            \multirow{2}{*}{\textbf{Assembly \#1}}  
                & $h = 100 \text{ cm},\, w=30 \text{ cm},\, d=30\text{ cm}$ & \\ 
                & $\text{mass} = 100 \text{ kg}$                            & \\ 
        \hline
            \multirow{2}{*}{\textbf{Assembly \#2}}  
                & $h = 10 \text{ cm},\, w=30 \text{ cm},\, d=30 \text{ cm}$    & $h = 90 \text{ cm},\, w=30 \text{ cm},\, d=30 \text{ cm}$ \\ 
                & $\text{mass} = 50\text{ kg}$                                        & $\text{mass} = 50\text{ kg}$ \\ 
        \hline \hline
    \end{tabular} 
\end{table}

\par Initial conditions for all test cases are assumed to be the following unless stated otherwise; zero translational velocity, zero rotational velocity, zero boom tip deflection, zero applied cable tension, and zero SIA, with $\vect{b}^3$ parallel to $\vect{s}$. The distance from the Sun for all cases is $1$ astronomical unit (au), or the mean radius of Earth's orbit. Amplitudes for the solar sail membrane quadrants, following the approach described in Section~\ref{sec:SolarSailModeling}, are specified for each test case. To assist with computational efficiency, artificial structural damping is added to the numerical simulation as described in~\cite{bunker2024modular} such that the damping ratio on all modes of the linear system do not exceed 1\%.

\subsection{Sail Shape Effect on SRP Torque}
\par
\begin{figure}[ht!]
    \centering
    \includegraphics[width=0.48\textwidth]{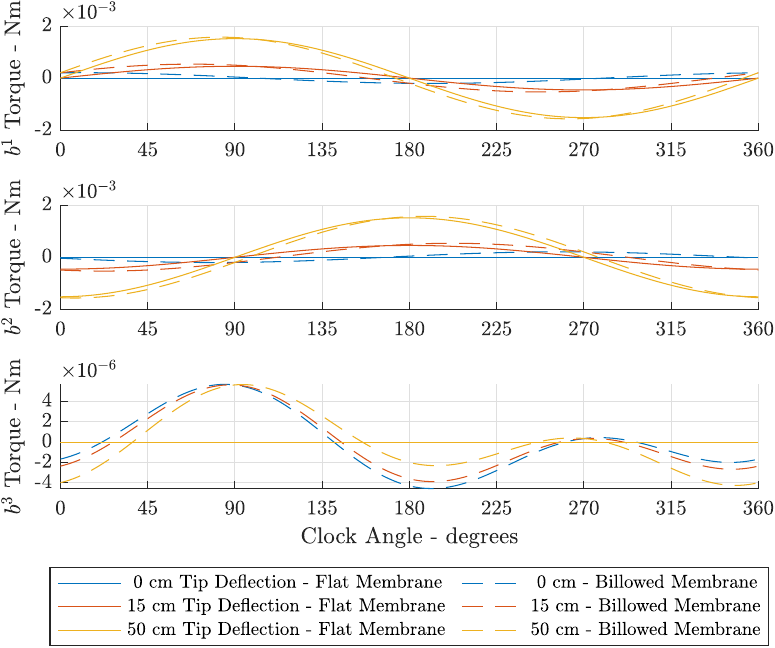}
    \caption{SRP torque generated by all six combinations of membranes and boom tip deflections across all clock angles at an SIA of $17^\circ$ and resolved in the body fixed frame, $\mathcal{F}_b$.}
    \label{fig: sail shape vs SRP torque TORQUES}
\end{figure}

\begin{figure}[ht!]
    \centering
    \subfloat[]{
        \includegraphics[width=0.48\textwidth]{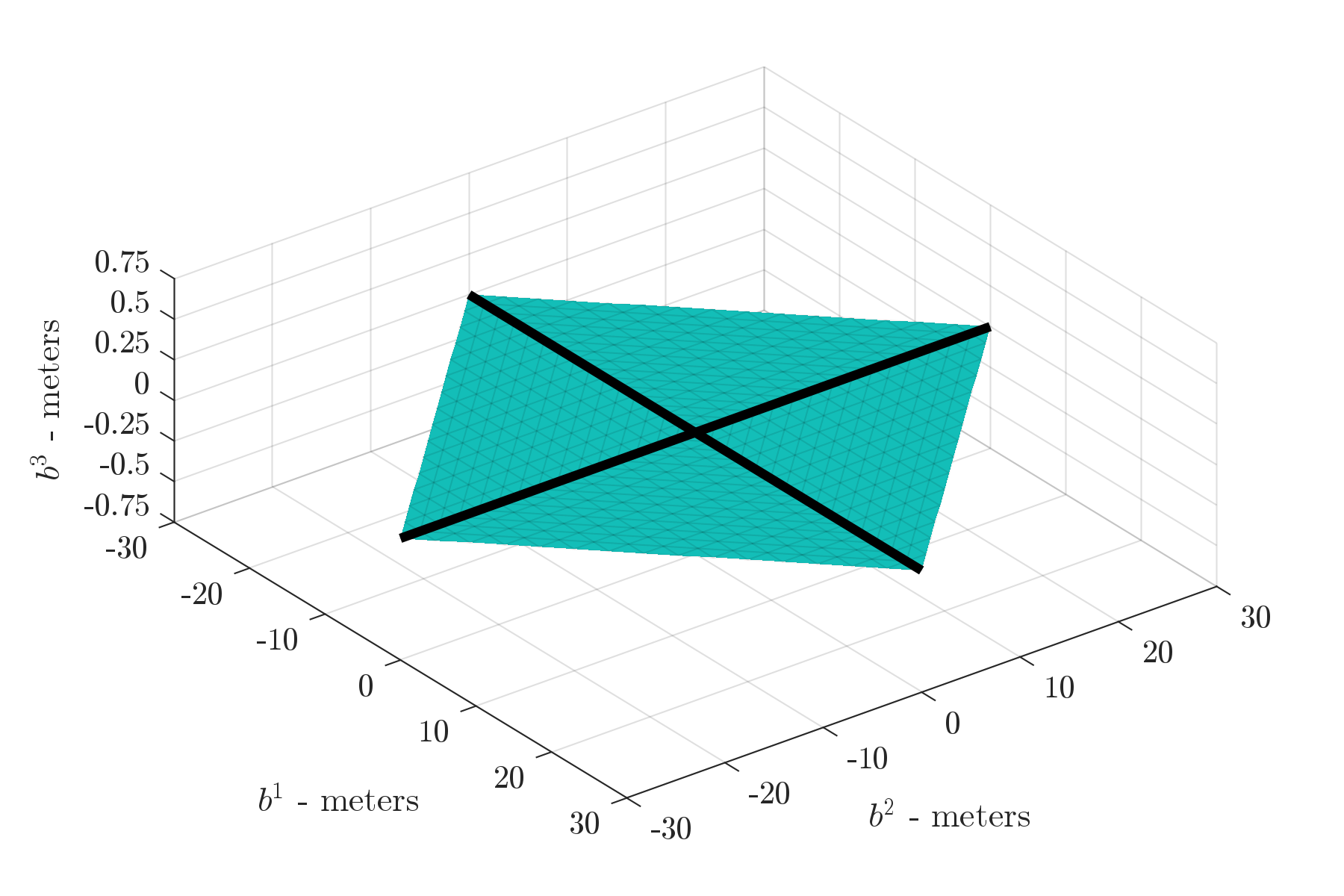}
    }
    \subfloat[]{
        \includegraphics[width=0.48\textwidth]{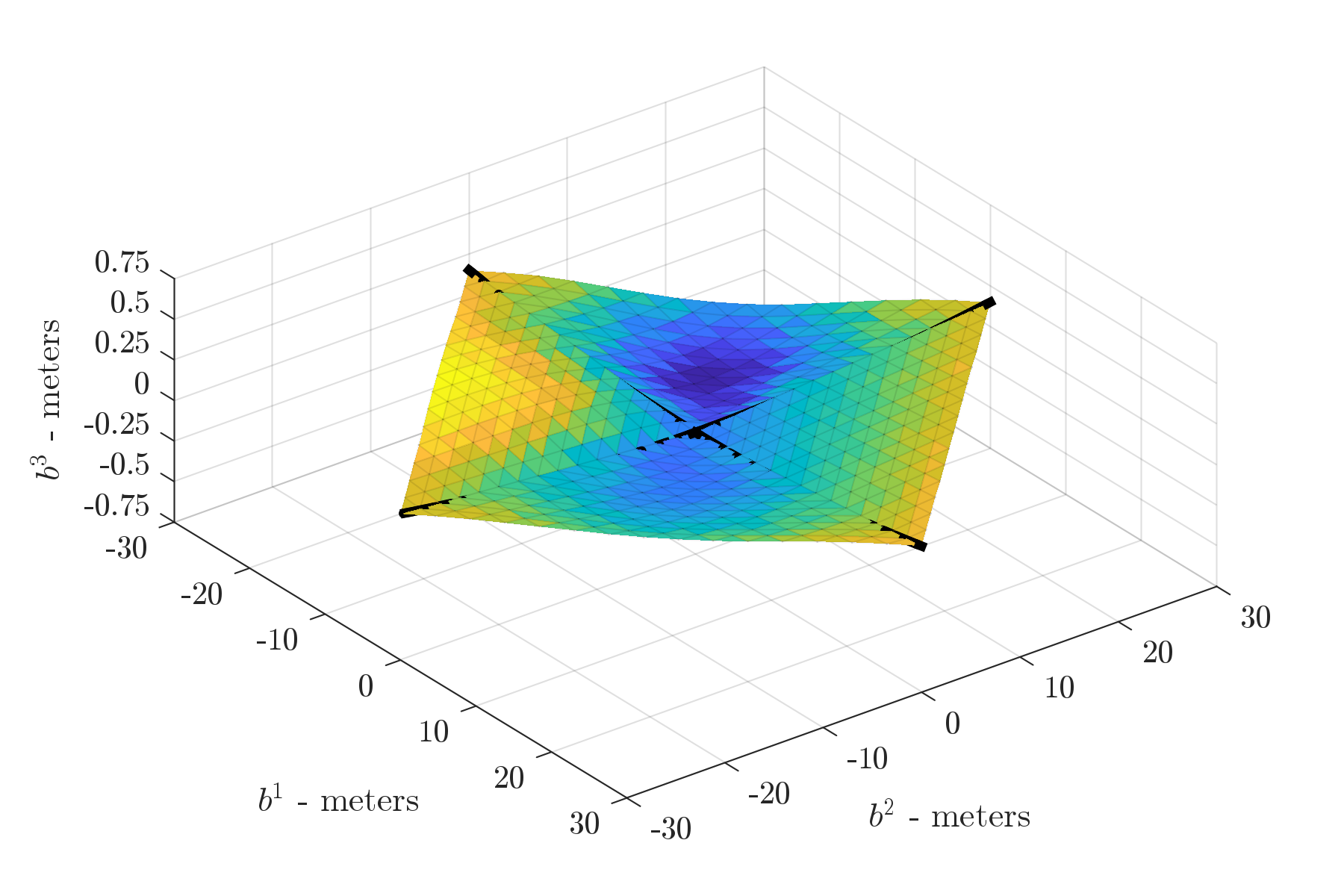}
    }
    \vspace{-8pt}
    \caption{(a) The flat membrane with zero boom tip deflection and (b) non-symmetric billowed membrane with symmetric boom tip deflection, representing the solar sail shapes considered to investigate the driver of SRP torque. Shading corresponds to membrane height along the $\vect{b}^3$ direction.}
    \label{fig: sail shape vs SRP torque SHAPES }
\end{figure}

 To investigate the effect of membrane shape and tip deflection on induced SRP torque, the SRP torque produced by two membrane shapes at three static boom tip deflections is calculated along all clock angles at a SIA of $17^\circ$. A flat membrane, with all quadrants having a $\Delta z_{max} = 0$, and a billowed membrane, with $\Delta z_{max} = \left[0 \text{ cm}, -15\text {cm}, 7.5\text {cm}, -7.5\text{ cm}\right]$ for the respective membrane quadrants, are considered to represent a perfectly flat sail membrane and a non-symmetric deformed, or ``billowed,'' sail membrane. Boom tip deflections of three magnitudes are considered: $0$ cm, $15$ cm, and $50$ cm to represent low, medium, and high boom tip deflections relative to the largest sail membrane deflection considered. One of the three boom tip deflections is applied to all four booms, and then the SRP torque is computed for both membranes applied to that static boom deflection. This is repeated for all three boom tip deflections. The SRP torque produced by these six resulting combinations of tip deflection and membranes are shown in Fig.~\ref{fig: sail shape vs SRP torque TORQUES}. The resulting solar sail shape from two of these membrane and tip deflection combinations are shown in Fig.~\ref{fig: sail shape vs SRP torque SHAPES }, demonstrating both membranes and zero versus non-zero tip deflection.

\par It can be seen that the change in tip deflection has a greater impact on the SRP torque than the change in membrane shape along the $\vect{b}^1$ and $\vect{b}^2$ directions, or the yaw and pitch directions respectively. In the $\vect{b}^3$, or roll, direction tip deflection does not change the SRP torque for the flat sail membrane. This is due to the flat membrane being a perfectly symmetric sail about the $\vect{b}^3$ direction, therefore it cannot produce roll torques. The SRP torque does vary with tip deflection under the billowed sail membrane. These tests were limited to a single SIA, two membrane shapes, and symmetric boom tip deflections. Nevertheless, these preliminary results indicate that a change in boom tip deflection can dominate the change in SRP torque compared to the effects of sail membrane billowing in the pitch and yaw directions, even in the presence of conservatively large membrane billowing. These results agree qualitatively with the SRP torque and sail shape relations outlined in \cite{gauvain2023solar}, which considered a sail membrane constrained along the entire boom length, rather than the tip and base as shown here.

\subsection{CABLESSail Tension Input}
\label{sec:CABLESSail Tension Input}
\par
\begin{figure}[hbt!]
    \centering
    \includegraphics[width=0.48\linewidth]{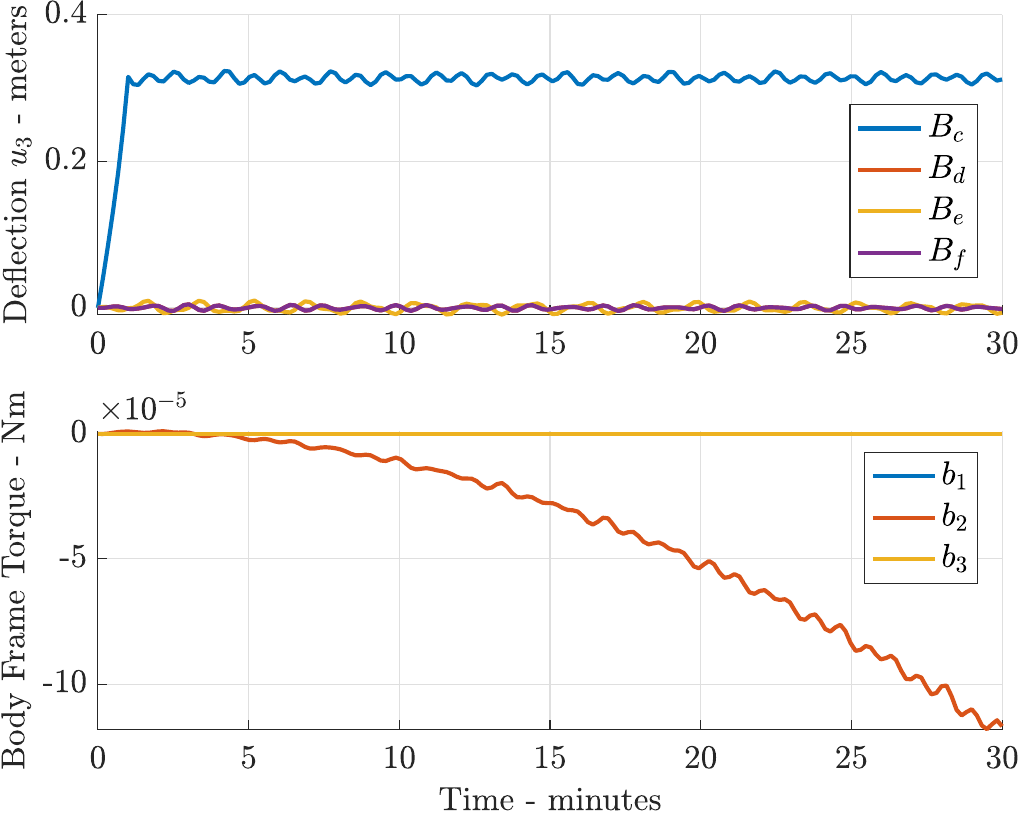}
    \caption{Time history of the out-of-plane boom tip deflections and resulting SRP body torques resolved in the solar sail bus frame, $\mathcal{F}_b$, subject to a one-minute linear ramp in tension followed by a constant tension in one cable.}
    \label{fig:CABLESSail tension input srp and tip deflection}
\end{figure}

 A numerical simulation of Assembly~\#1 is conducted with a flat membrane for each quadrant ($\Delta z_{max} = 0$). All cable tensions are kept zero except for a cable on the boom along the positive $\vect{b}^1$ direction, or boom 1 following the convention of~Fig.~\ref{fig: sail membrane coordiantes}, where a linear ramp in tension is applied across the first minute of the simulation, ending at a value of 4.0 N. The final tension is held constant for 29 additional minutes following the linear ramp, where the simulation is then concluded after a total elapsed time of 30 minutes. Time histories of the out-of-plane boom tip deflections and the resulting SRP torque applied to the assembly, resolved in $\mathcal{F}_b$, are shown in Fig.~\ref{fig:CABLESSail tension input srp and tip deflection}.

\par The boom tip deflection is observed to have nearly a quadratic dependence on cable tension as the linear ramp is applied. This agrees with equilibrium boom tip deflections observed under constant cable tensioning for similar dynamic boom models~\cite{Lee2026ActaAstronautica}. It can be seen that oscillations are present in all boom tip deflections, indicative of vibrations in the whole structure once a non-zero open-loop cable tension is applied without consideration for the resulting dynamic response. These vibrations are carried forward into the SRP torques due to the boom tip sail attachment points. Vibrations are expected with actuation of flexible structures, but are generally undesirable for momentum management torques, so active feedback control of the structure will be required to eliminate this behavior. Torque of this magnitude is comparable to other momentum management technologies for solar sails of this size~\cite{inness2023momentum}.

\subsection{AMT vs CABLESSail Attitude Change}
\begin{figure}[ht!]
    \centering
    \includegraphics[width=0.48\linewidth]{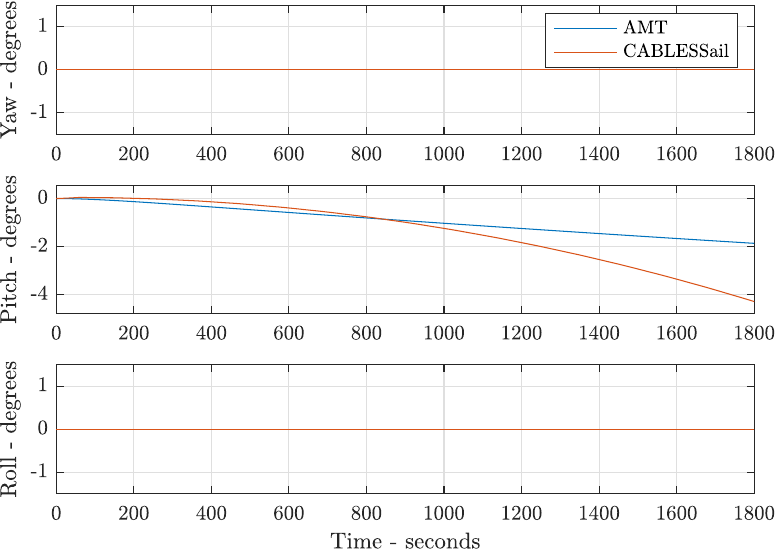}
    \caption{Euler angle time histories of (a) Assembly \#1 generating SRP torque through cable tensioning and deflection of one boom and (b) Assembly \#2 with a static AMT actuation of $30$~cm spacecraft bus offset.}
    \label{fig: AMT vs CABLESSail Attitude Change}
\end{figure}

Two numerical simulations are conducted of the Assembly~\#1 and Assembly~\#2 models individually with nominal initial conditions. Assembly \#1 in the first simulation begins with zero cable tensions, then an identical linear ramp in tension to that described in Section~\ref{sec:CABLESSail Tension Input} is applied to a single cable, bending boom $1$ out of the sail plane towards the Sun-facing direction. Assembly \#2 in the second simulation utilizes a constant offset between the spacecraft bus and the solar sail bus, such that the center of mass of the spacecraft bus is $30$ cm offset from the center of mass of the spacecraft bus in the $\vect{b}^1$ direction, which is consistent with the maximum achievable offset by Solar Cruiser's AMT~\cite{johnson2020solar,inness2023momentum}. No tension is applied to any cables within the Assembly~\#2 model for the duration of the simulation. Both models are simulated for 30 minutes, and the attitude time history is shown in Fig. \ref{fig: AMT vs CABLESSail Attitude Change} as Euler angles. It can be seen that the CABLESSail boom actuation of Assembly~\#1 meets and then exceeds the pitch angle change of the AMT in Assembly~\#2. The attitude profile of the Assembly~\#1 simulation shows an increasing rate of change in the pitch angle, while the Assembly~\#2 simulation shows a roughly constant pitch rate across this testing window. This initial test provides confidence that a reasonable tensioning of CABLESSail's actuating cables could be an effective means to generate usable SRP control torques in comparison to an AMT.

\subsection{CABLESSail SRP Torque Change Capability}
\par A Monte Carlo test is conducted with the static model of Assembly~\#1 to demonstrate the capability of CABLESSail to alter the SRP torque produced from coordinated boom deflections under unknown sail membrane shape conditions. Two specific boom deflection combinations are identified for examination: one likely to produce torques in the pitch-yaw directions by deforming one boom and one likely to produce torques in the roll direction by deforming all booms. These two combinations, or maneuvers, are detailed in their following subsections. Each begins with zero initial boom tip deflection then applies a $50 \text{ cm }$ magnitude cable-actuated tip deflection to the identified booms in either the positive or negative direction. To represent the membrane shape uncertainty, one hundred random sail membrane shapes are generated by independently sampling membrane function amplitudes, $\Delta z_{max}$, for each quadrant from a uniform distribution on the range $(-15.0 \text{ cm, } 15.0 \text{ cm})$ Exhaustively testing these maneuvers under all SIAs and clock angles is outside the scope of this note, as this is a preliminary investigation into the capabilities of CABLESSail. Thus, each maneuver is only tested under one SIA and clock angle pair. Both are tested at the maximum SIA considered for Solar Cruiser's attitude control analysis, $17^\circ$, and each are tested at a clock angle where reasonable torque change values are expected, $0^\circ$ and $45^\circ$ for maneuver one and maneuver two respectively.

\par To investigate the effect of boom tip deflection magnitude on SRP torques, each maneuver is progressively applied to the solar sail and the $100$ random membrane shapes. The induced SRP torque is recorded as increasing boom tip deflections are applied to each membrane case, sweeping through tip deflections from $0$ cm to $50$ cm in the respective directions for each boom of each maneuver. To assess the ability of each maneuver to alter the SRP torque, the difference in SRP torque between the fully actuated maneuver and initial tip deflection of $0$ cm is computed for each of the $100$ membranes. This change in SRP torque is shown through two sets of binned histograms; one depicting the total change in SRP along each axis, and the other depicting the percent change along each axis.

\subsubsection{Pitch - Yaw Torque Generation}
\par
\begin{figure}[hbt!]
    \centering
    \includegraphics[width=0.48\linewidth]{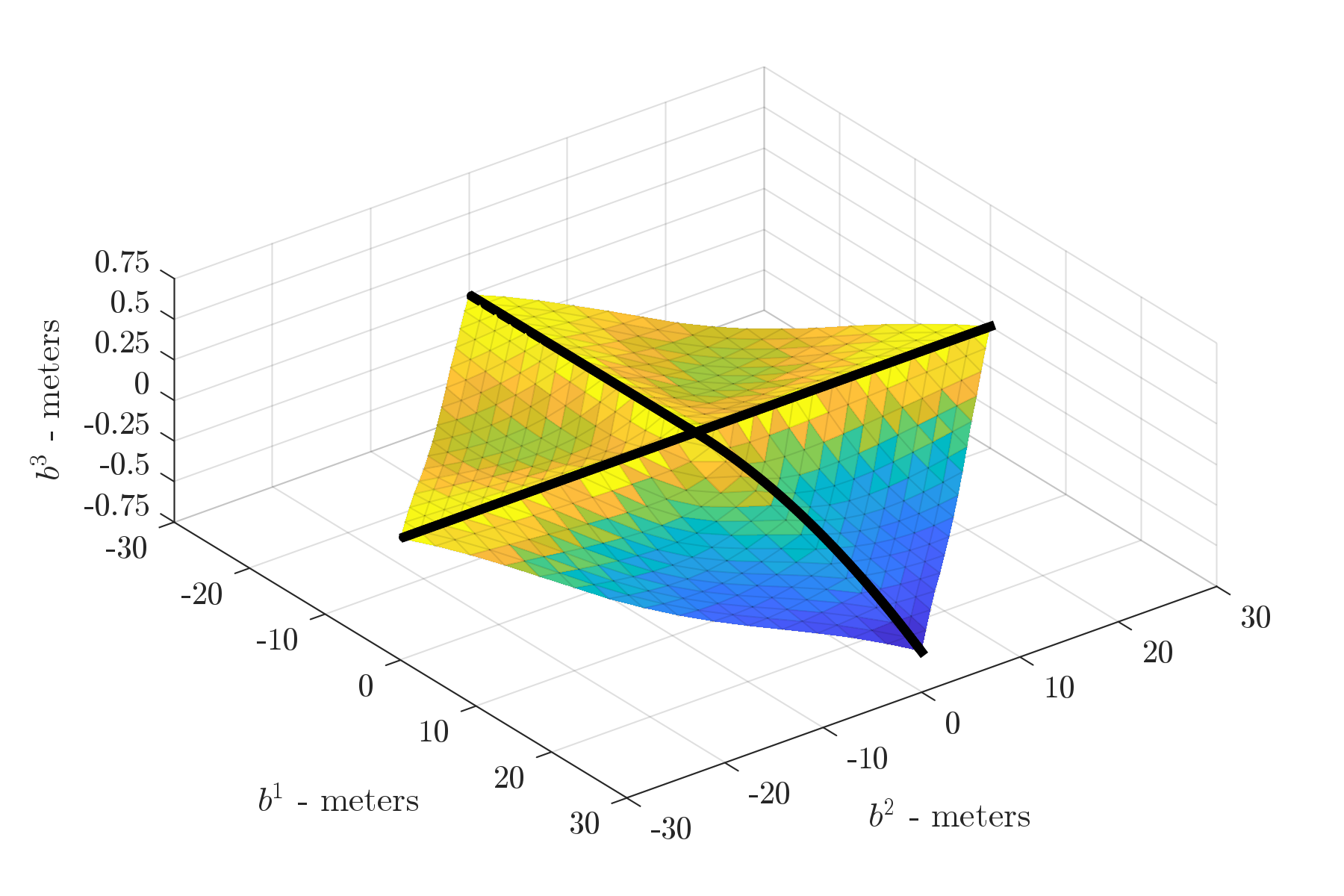}
    \caption{Static CABLESSail boom deformations under the first boom maneuver, where the boom along the positive $\vect{b}^1$ direction is deflected by $-50$ cm. Shading corresponds to membrane height along the $\vect{b}^3$ direction.}
    \label{fig:Capability Pitch Yaw Sail shape}
\end{figure}

\begin{figure}[ht!]
    \centering
    \includegraphics[width=0.48\textwidth]{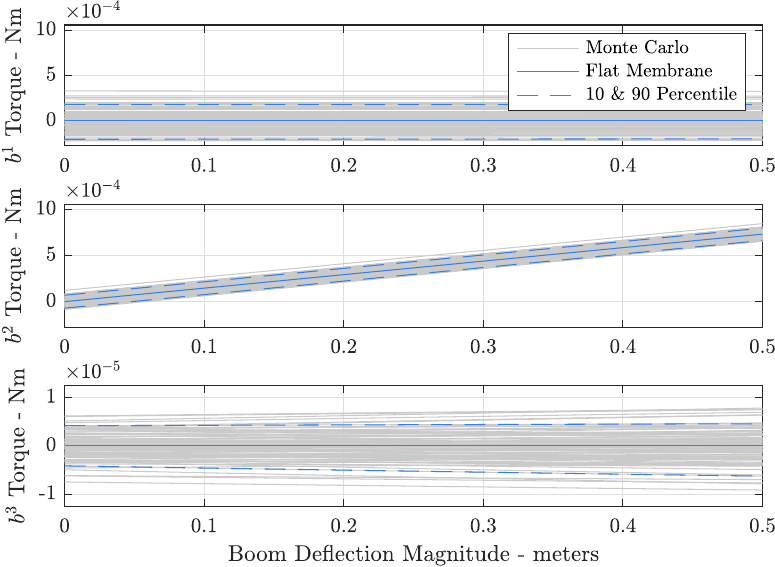}
    \caption{SRP torque generated by 100 randomly deformed solar sail membrane shapes at a clock angle of $0^\circ$ and SIA of $17^\circ$ through progressive application of maneuver 1. }
    \label{fig:Capability Pitch Yaw sweeping maneuver}
\end{figure}

  The first boom-maneuver consists of deforming boom $1$ away from the sun, the negative direction, with all other booms left at the nominal zero deflection. This deformed sail structure is shown in Fig. \ref{fig:Capability Pitch Yaw Sail shape}. The SRP torque resolved in $\mathcal{F}_b$ as the maneuver is progressively applied to all one hundred randomly generated membranes at an SIA of $17^\circ$ and a clock angle of $0^\circ$ is show in Figure~\ref{fig:Capability Pitch Yaw sweeping maneuver}. To quantify the dispersion of the SRP torque across all membranes, the flat membrane ($\Delta z_{max}=0$) and the membrane cases with the 10th and 90th percentile of initial SRP torque at zero boom deflection along each direction are highlighted. For this maneuver the yaw ($\vect{b}^1$) and roll ($\vect{b}^3$) torques see minimal change across all membrane cases as the maneuver is applied. Meanwhile, the pitch ($\vect{b}^2$) torque increases in roughly a linear fashion across all membranes, with the total torque change from the beginning to the end of the maneuver exceeding the variation in torque between the membranes. With the membrane deflection and tip deflections chosen for this example, it can be seen that the tip deflection is the primary driver of SRP torque change, dominating any affects from the randomized sail membranes. A larger $\Delta z_{max}$ increases the variation of torques between the randomized sail membranes. The choice of a $\Delta z_{max}$ in this study is larger than past mission analyses, demonstrating preliminary CABLESSail performance under conservative conditions. The trends of SRP torque under boom actuation justify the use of a lower fidelity membrane model for this preliminary study since any change in SRP torque from a higher-fidelity model would be dwarfed by the SRP torque change across the maneuver.

\begin{figure}[hbt!]
    \centering
    \subfloat[]{
        \includegraphics[width=0.48\linewidth]{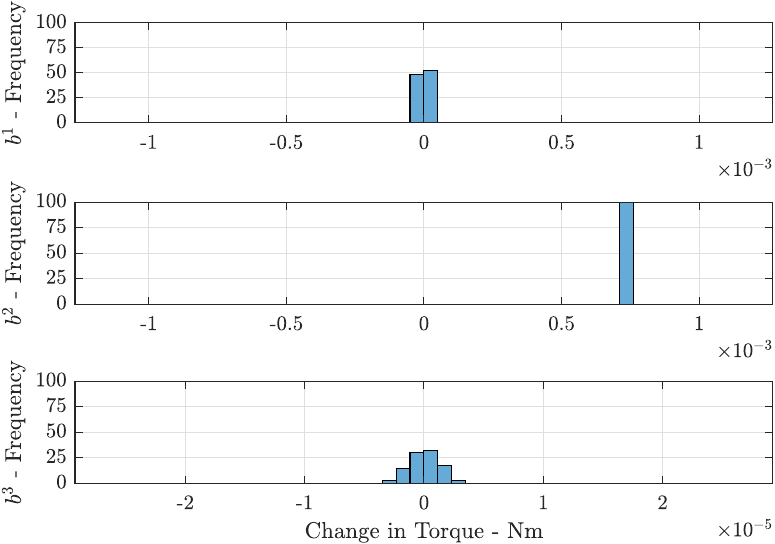}  
        \label{fig:12a}
   }
    \subfloat[]{
         \includegraphics[width=0.45\linewidth]{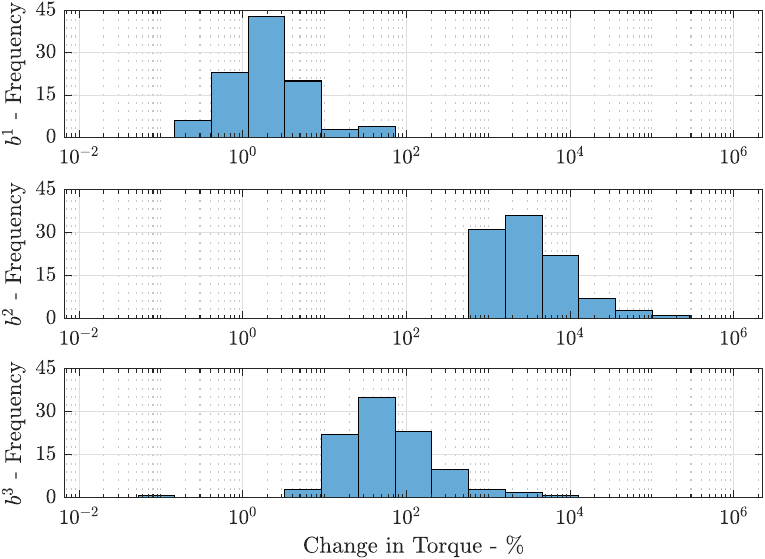}  
         \label{fig:12b}
    }
    \vspace{-8pt}
    \caption{Change in SRP torque by applying maneuver one to $100$ random membrane shapes, reported in total change, (a), and percent change, (b), along the directions of $\mathcal{F}_b$. }
    \label{fig:Capability Pitch Yaw Sail histogram}
\end{figure}

Change in SRP torque along the $\mathcal{F}_b$ directions for all one hundred membrane cases at the prescribed SIA and clock angle is shown in Fig. \ref{fig:Capability Pitch Yaw Sail histogram} as a binned histogram. For all cases there is a consistent positive change in torque along the pitch direction, with random dispersion along yaw and roll centered on zero. The percent change in pitch torque is the largest amongst all directions, exceeding the percent change of nearly all cases in yaw and roll by an order of magnitude.

\par Given the reliable generation of a relatively large pitch torque with minimal yaw torque and roll torque, this particular maneuver shows great promise in situations where a pitch torque is required for attitude control or momentum management. As the boom structure of the solar sail is symmetric, and the membrane shapes are random, this result can be extrapolated for any clock angle aligning with a boom ($0^\circ$, $90^\circ$, $180^\circ$, $270^\circ$) and deforming the appropriate boom according to this maneuver. This result shows promising evidence of CABLESSail's capability to counteract SRP disturbance torques, with the average pitch torque change of all test cases meeting the maximum pitch-yaw disturbance torque expected at this SIA for Solar Cruiser~\cite{gauvain2023solar}.

\subsubsection{Roll Torque Generation}
\par
\begin{figure}[bht!]
    \centering
    \includegraphics[width=0.48\linewidth]{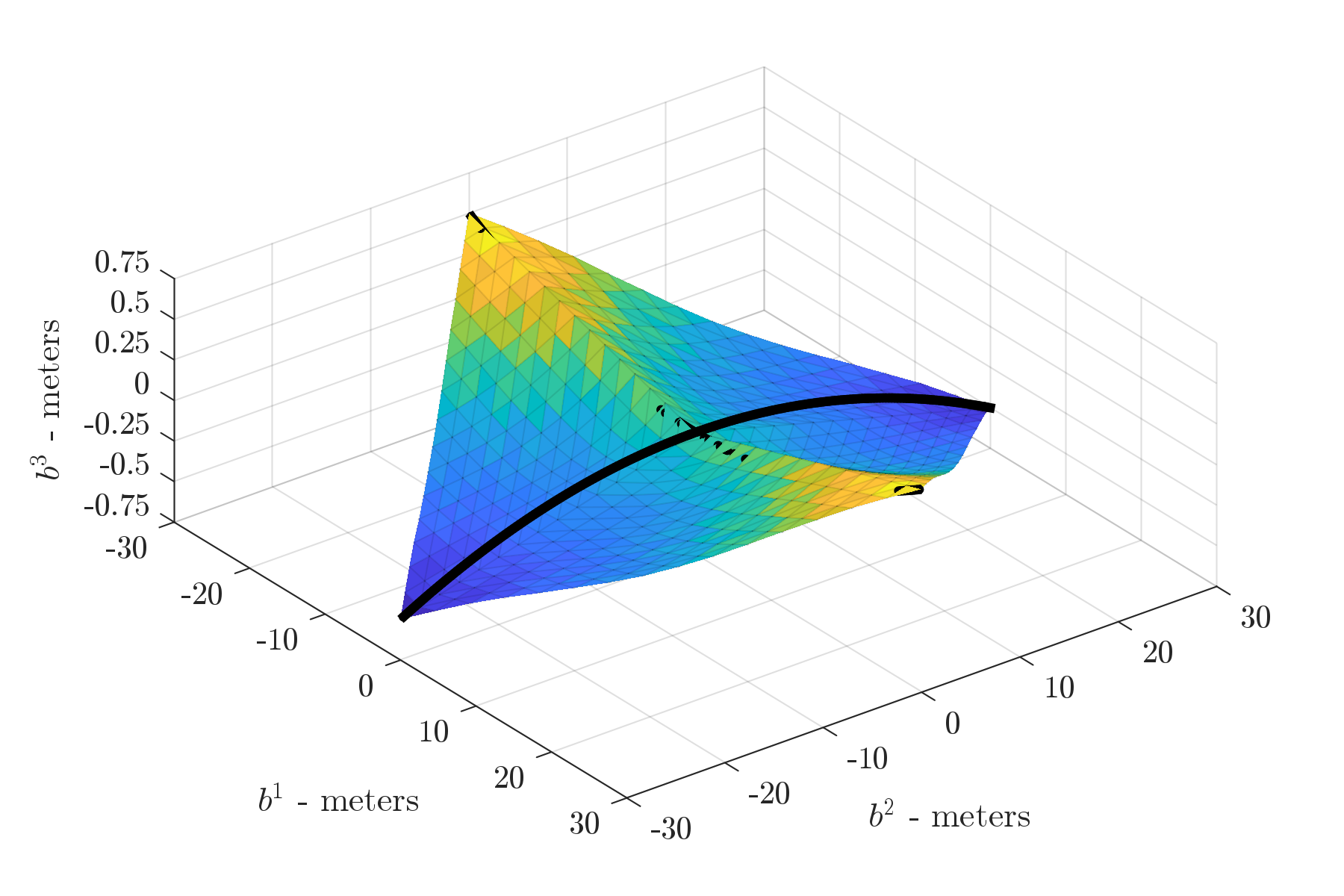}
    \caption{ Static boom deformations for the second boom maneuver. Booms one and three are deformed by $50$ cm, booms two and four are deformed by $-50$ cm. Shading corresponds to membrane height along the $\vect{b}^3$ direction.}
    \label{fig:Capability Roll Sail shape}
\end{figure}

\begin{figure}[bht!]
    \centering
    \includegraphics[width=0.48\textwidth]{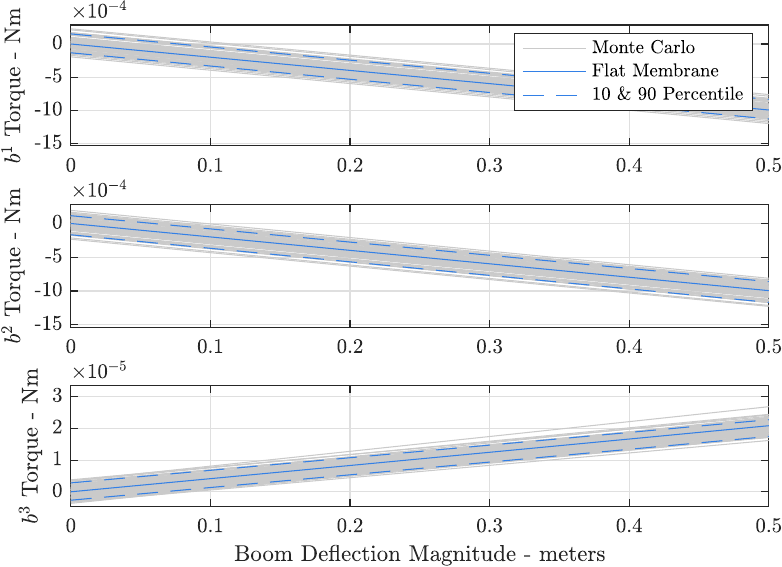}
    \caption{SRP torque generated by 100 randomly deformed solar sail membrane shapes at a clock angle of $0^\circ$ and SIA of $17^\circ$ through progressive application of maneuver 2. }
    \label{fig:Capability Roll sweeping maneuver}
\end{figure}

\begin{figure}[hbt!]
    \centering
    \subfloat[]{
        \includegraphics[width=0.48\linewidth]{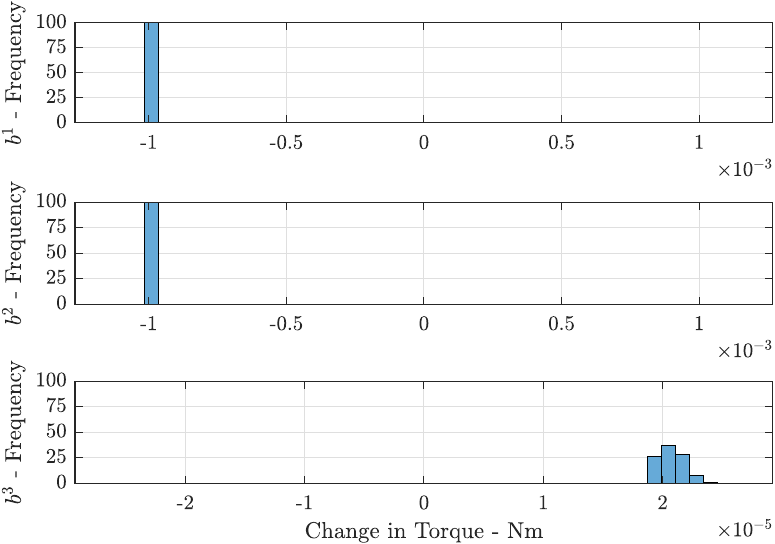}  
        \label{fig:15a}
   }
    \subfloat[]{
        \includegraphics[width=0.45\linewidth]{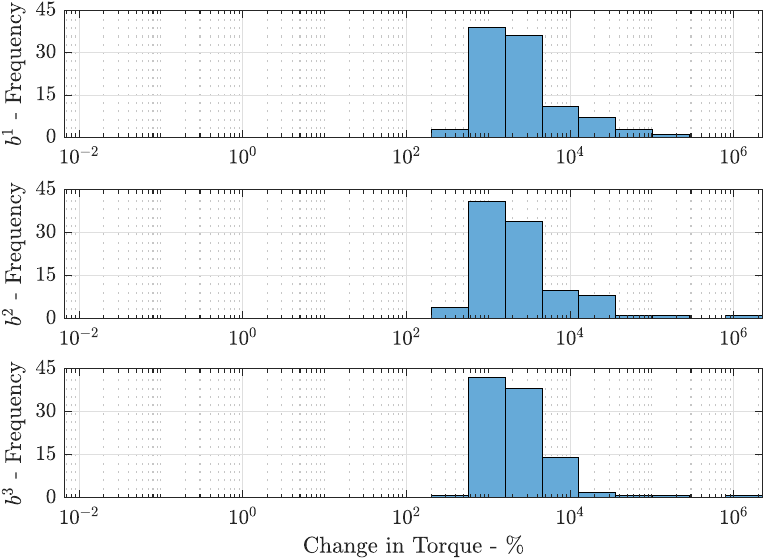}  
        \label{fig:15b}
    }
    \vspace{-8pt}
    \caption{ Change in SRP torque by applying maneuver two to $100$ random membrane shapes, reported in total change, (a), and percent change, (b), along the directions of $\mathcal{F}_b$.}
    \label{fig:Capability Roll Sail histogram}
\end{figure}

 The second boom-maneuver deflects booms 1 and 3 in the positive direction, and booms 2 and 4 in the negative direction. Fig.~\ref{fig:Capability Roll Sail shape} shows the resulting deformed sail structure. The SRP torque resolved in $\mathcal{F}_b$ as the maneuver is progressively applied to all one hundred membranes at a clock angle of $45^\circ$ and SIA of $17^\circ$ is shown in Fig.~\ref{fig:Capability Roll sweeping maneuver}. The roll direction has a roughly linear increase in torque across all membranes as the maneuver is applied. The remaining two directions, pitch and yaw, also have a linear change in torque. The total change in torque from the beginning to end of the maneuver along each direction for each membrane is shown in Fig.~\ref{fig:Capability Roll Sail histogram} as a binned histogram. The roll direction shows a consistent positive change in torque with some dispersion between membrane cases, while the pitch and yaw directions both show a consistent negative change in torque of similar amounts amongst all membrane cases. 

Although this maneuver generates torques in all directions, it is noteworthy that a reliable large change in roll torque is generated.The maximum change in torque generated in the roll axis is on the order of $10^{-5}$~Nm, which is an equal order of magnitude to the torque capability of Solar Cruiser's reflectivity control devices~\cite{heaton2023reflectivity} and the largest roll disturbance torque predicted for Solar Cruiser~\cite{gauvain2023solar}. Roll torques are notoriously challenging to generate with the actuators available to solar sails while undesired pitch and yaw torques are much simpler to address~\cite{inness2023momentum}, which makes this actuation maneuver potentially useful. Future work will investigate a control allocation method to determine the necessary boom deflections and cable tensions to generate desired torques at any clock angle and SIA. 

\section{Conclusions}
\label{sec:Conclusion}

\par Through open-loop cable tensioning, CABLESSail surpassed the attitude slewing performance of a fully actuated AMT in a preliminary testing scenario. Dynamic structural response to open-loop cable tensioning indicated the need for robust feedback control of boom deflections. CABLESSail demonstrated large roll and pitch torque generation at a subset of SIA and clock angles under solar sail membrane shape uncertainty that exceed the membrane deflection considered for the Solar Cruiser attitude control analyses, with boom tip deflections driving the change in SRP torque over sail membrane shape. The pitch and yaw torques generated met the Solar Cruiser disturbance torque requirements, and roll torques were generated with residual pitch and yaw torques.

\pagebreak
\section*{Appendix: Component Dynamic Modeling}
\label{sec:Appendix Component Dynamic Modeling}



\par The individual components are modeled by first identifying the generalized coordinates and velocities, then deriving the component kinetic and potential energies. The generalized forces acting on the components are derived and Lagrange's equation is used to derive the constrained component equations of motion. Finally, the first instance of the null-space method is used to remove the Lagrange multipliers enforcing the attitude parameterization constraint. All components are assumed to be of uniform density, which is not a requirement of the method, but a convenient assumption for the analyses to follow.

\section{Additional Preliminaries: Physical Vectors}
The physical vector of the position of point $p$ relative to point $w$ is $\vect{r}^{pw}$. The velocity of point $p$ relative to point $w$ with the time derivative taken with respect to $\mathcal{F}_a$ is, $\vect{v}^{pw/a} = \vect{r}^{pa^{\cdot a}}$. The angular velocity of $\mathcal{F}_b$ relative to $\mathcal{F}_a$ is $\vect{\omega}^{ba}$. Position, velocity, angular velocity, and all other physical vectors can be resolved in any reference frame. For example, $\mathbf{r}_a^{pw}$, $\mathbf{v}_a^{pw/a}$, and $\mbs{\omega}_a^{ba}$, represent physical vectors resolved in $\mathcal{F}_a$.

\section{Attitude Parameterization for Each Component}

\par The attitude of all component bodies, $B_i$, where $i \in \{b, c, d, e, f, g\}$, is defined as the attitude between the respective component body frame, $\mathcal{F}_{i}$, affixed to each body and the inertial frame, $\mathcal{F}_{a}$, as expressed through the direction cosine matrix, $\mbf{C}_{ia}$.  All DCMs are vectorized into a generalized coordinate as column vector $\mbf{p}^{ia}$,  where 
$\mathbf{C}_{ia}^{\trans} = 
    \Big[
        \mbf{p}_1^{ia} \,\,\, \mbf{p}_2^{ia}  \,\,\, \mbf{p}_3^{ia} 
    \Big]$ 
and 
$\mathbf{p}^{ia^\trans} = 
    \Big[
        \mbf{p}_1^{ia^\trans} \,\,\, \mbf{p}_2^{ia^\trans} \,\,\, \mbf{p}_3^{ia^\trans} 
    \Big]$. 

\par The rates of $\mbf{p}^{ia}$ are related to the angular velocity of $\mathcal{F}_i$ relative to $\mathcal{F}_a$ resolved in $\mathcal{F}_i$ by
\begin{align} \label{S_pdot_to_omega}
    \dot{\mbf{p}}^{ia} &= \mbs{\Gamma}^{ia}_i(\mbf{p}^{ia})\mbs{\omega}^{ia}_i ,\\
     \mbs{\omega}_i^{ia} &= \mbf{S}(\mbf{p}^{ia})\dot{\mbf{p}}^{ia} ,
\end{align}
where
\begin{equation*}
    \mbf{S}(\mbf{p}^{ia}) = \mbf{S}_i^{ia} = 
    \begin{bmatrix}
        \mbf{0}                 & \mbf{p}_3^{ia^\trans}    & \mbf{0} \\
        \mbf{0}                 & \mbf{0}                  & \mbf{p}_1^{ia^\trans} \\
        \mbf{p}_2^{ia^\trans}   & \mbf{0}                  & \mbf{0}
    \end{bmatrix}
    , \quad
    \mbs{\Gamma}(\mbf{p}^{ia}) = \mbs{\Gamma}_i^{ia} = 
    \begin{bmatrix}
        \mbf{0}         & \mbf{0}           & \mbf{p}_2^{ia} \\
        \mbf{p}_3^{ia}  & \mbf{0}           & \mbf{0} \\
        \mbf{0}         & \mbf{p}_1^{ia}    & \mbf{0}
    \end{bmatrix} ,
\end{equation*} 
and $\mbf{S}^{ia}_i \mbs{\Gamma}^{ia}_i = \mbf{1}$~\cite{de_ruiter_SO3parameterixation_identities_2014}.

\section{Component Generalized Coordinates \& Generalized Velocities}

\par The generalized coordinates of all components include a position vector relative to an inertial point and a vectorized DCM relative to an inertial frame, and the flexible booms utilize elastic coordinates to describe the deformation state. Generalized velocities of all components consist of the respective time derivatives of these coordinates. 

\subsection{Solar Sail Bus}
\par
\begin{figure}[hbt!]
    \centering
    \includegraphics[width=0.35\linewidth]{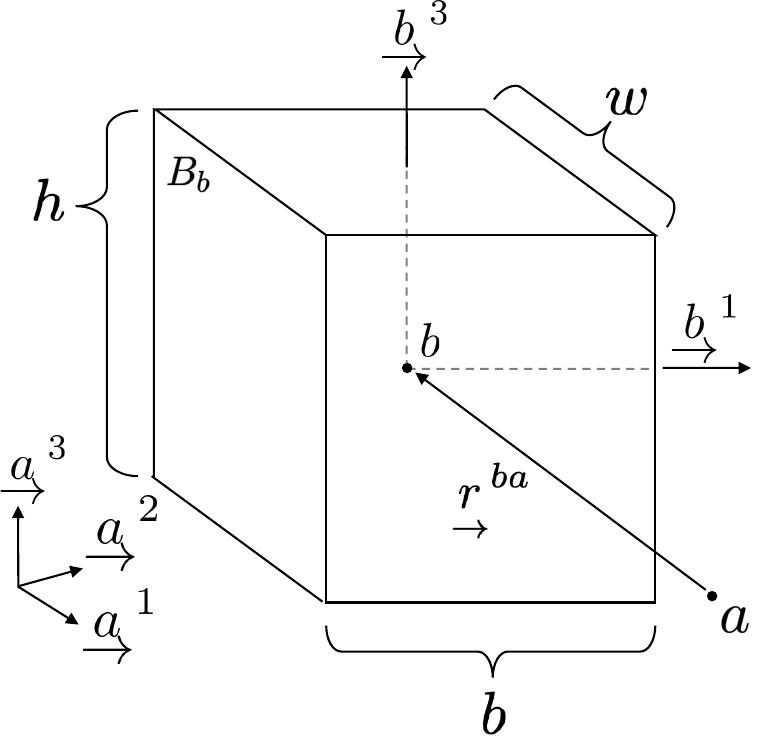}
    \caption{Solar sail bus component, $B_b$, with body-fixed reference frame, $\mathcal{F}_b$, center of mass, $b$, position relative to the inertial point $a$, $\protect\vect{r}^{ba}$, and base, $b$, height, $h$, and width, $w$, dimensions labeled. }
    \label{fig:solar sail bus diagram}
\end{figure}

 The solar sail bus, body $B_b$, is modeled as a continuous rigid rectangular prism with dimensions $h$, $w$, and $d$, as shown in Fig.~\ref{fig:solar sail bus diagram}. Reference frame $\mathcal{F}_{b}$ is affixed to the body. The $\vect{b}_1$, $\vect{b}_2$, and $\vect{b}_3$ directions are referred to throughout the note as yaw, pitch, and roll directions, respectively, for convenience. The plane defined by $\vect{b}_1$ and $\vect{b}_2$ is the solar sail plane. The generalized coordinates describing the state of $B_b$ are $\mathbf{q}^{B_b^\trans} = \Big[
        \mathbf{r}_a^{ba^\trans} \,\,\, \mathbf{p}^{ba^\trans}
    \Big]$,
where $\mathbf{r}_a^{ba}$ is the position of the center of mass of the body relative to an unforced particle $a$ resolved in the inertial frame $\mathcal{F}_{a}$, and where $\mathbf{p}^{ba}$ is the vectorized DCM $\mbf{C}_{ba}$. The generalized velocities for the unconstrained spacecraft hub component $B_{b}$ are $\mbs{\nu}^{B_b^\trans} =
\Big[
    \dot{\mbf{r}}_a^{ba^\trans} \,\,\,
    \mbs{\omega}_b^{ba^\trans}
\Big]$, 
where $\mbs{\omega}_b^{ba}$ is the angular velocity vector of body frame $\mathcal{F}_{b}$ relative to $\mathcal{F}_{a}$ resolved in $\mathcal{F}_{b}$. These generalized velocities are related to the generalized coordinates through the mappings 
\begin{align}\label{eq:HubKinMap1}
    \dot{\mbf{q}}^{B_b} &= \mbs{\Gamma}_{B_b}\mbs{\nu}^{B_b} \\
    \mbs{\nu}^{B_b} &= \mbf{S}_{B_b} \dot{\mbf{q}}^{B_b}    
\end{align}
where $\mbs{\Gamma}_{B_b} = \text{diag}\{ \mbf{1}, \;  \mbs{\Gamma}^{ba}_b(\mbf{p}^{ba}) \}$, $\mbf{S}_{B_b} = \text{diag}\{ \mbf{1}, \;  \mbf{S}^{ba}_b(\mbf{p}^{ba})\}$, and $\mbf{S}_{B_b}\mbs{\Gamma}_{B_b} = \mbf{1}$. The dimensions and inertial properties of this component are varied in the numerical tests cases and are specified in Section~\ref{sec:Numerical simulation}.

\subsection{Spacecraft Bus}
\par
\begin{figure}[hbt!]
    \centering
    \includegraphics[width=0.4\linewidth]{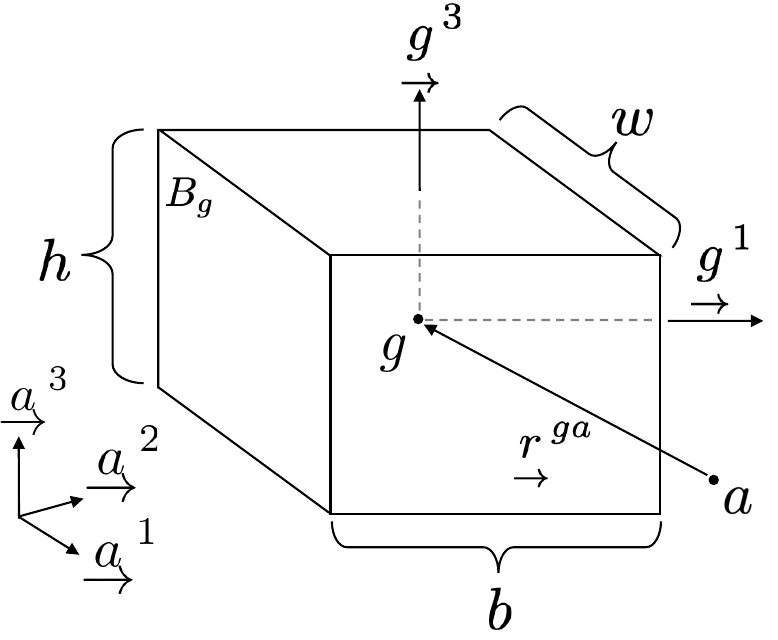}
    \caption{Spacecraft bus component, $B_g$, with body-fixed reference frame, $\mathcal{F}_g$, center of mass, $g$, position relative to the inertial point $a$, $\protect\vect{r}^{ga}$, and base, $b$, height, $h$, and width, $w$, dimensions labeled. }
    \label{fig:spacecrat bus diagram}
\end{figure}

 The spacecraft bus, body $B_g$, is modeled as a solid rectangular prism with dimensions $h$, $w$, and $d$, with body fixed reference frame $\mathcal{F}_g$ as show in Fig.~\ref{fig:spacecrat bus diagram}. The offset between the solar sail bus and spacecraft bus achieved by the AMT is held constant for all test cases in the work presented here, thus the AMT generalized coordinates and the relation to the generalized velocities follow similarly to the spacecraft bus as
$\mathbf{q}^{B_g^\trans} = \Big[
        \mathbf{r}_a^{ga^\trans} \,\,\, \mathbf{p}^{ga^\trans}
\Big]$,
$\mbs{\nu}^{B_g^\trans} =
\Big[
    \dot{\mbf{r}}_a^{ga^\trans} \,\,\,
    \mbs{\omega}_b^{ga^\trans}
\Big]$,
$\dot{\mbf{q}}^{B_g} = \mbs{\Gamma}_{B_g}\mbs{\nu}^{B_g}$,
and $\mbs{\nu}^{B_g} = \mbf{S}_{B_g} \dot{\mbf{q}}^{B_g}$. The inertial quantities and dimensions are noted in Section~\ref{sec:Numerical simulation}.

\subsection{CABLESSail Booms}
\par
\begin{figure}[t!]
    \centering
    \includegraphics[width=0.75\linewidth]{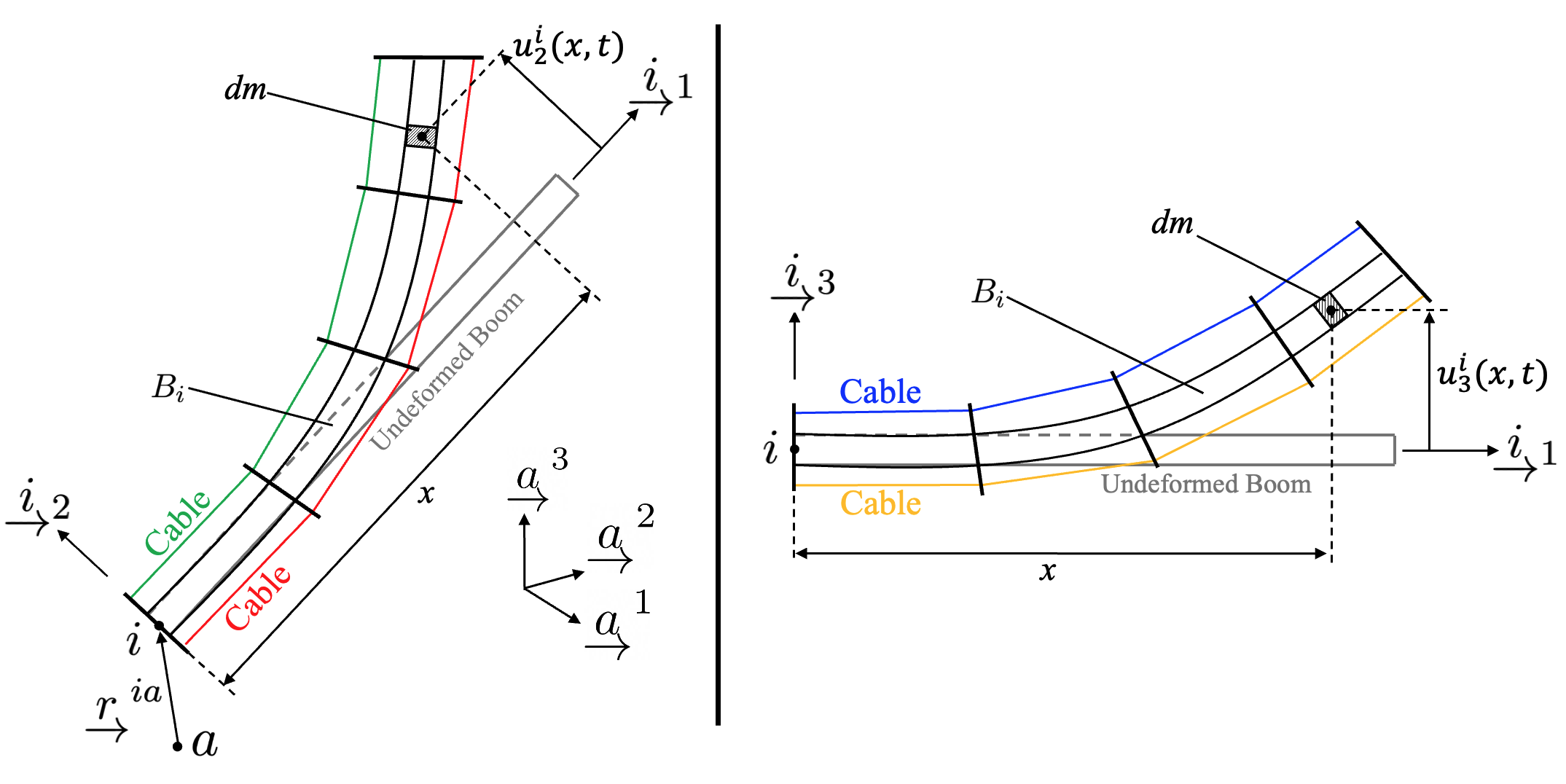}
    \caption{Boom component, $B_i$, with deflection $u_1^i(x,t)$ and $u_2^i(x,t)$ in two perpendicular planes shown with the actuation cables.}
    \label{fig:CABLESSail boom diagram}
\end{figure}

 Each flexible boom is modeled as a cantilevered flexible cylinder referred to as body $B_{i}, \text{ where } i \in \left\{ c, d, e, f\right\}$. Reference frame $\mathcal{F}_{i}$ is affixed to the base of each boom, as shown in Fig. \ref{fig:CABLESSail boom diagram}. The total deflection is modeled as a superposition of bending in two orthogonal planes. For $B_i$ these planes are the $\vect{i}_1-\vect{i}_2$ plane and $\vect{i}_1-\vect{i}_3$ planes. The deflection is modeled with the assumed modes method \cite{Rao2019}, which involves the deflection at point $x$ along the boom $B_i$ at time $t$ to be computed as
\begin{equation}\label{eq: deflection definition}
    \mathbf{u}^{i}(x,t) = \bbm u_2^i(x,t) \\ u_3^i(x,t)\ebm = \mathbf{\Psi}(x) \mathbf{q}^{\epsilon i}(t) ,
\end{equation}
where $u_2^i(x,t)$ and $u_3^i(x,t)$ are the elastic deformations at a distance $x$ aong the boom (i.e., in the $\vect{i}_1$ axis) in the $\vect{i}_2$ and $\vect{i}_3$ axes, respectively;
\begin{equation*}
    \mathbf{\Psi}(x) = \bbm \mbs{\Psi}_{u_2}(x) \\ \mbs{\Psi}_{u_2}(x) \ebm = \begin{bmatrix}
        \Psi_1 (x) & \Psi_2 (x) & \dots & \Psi_n (x) & 0 & 0 & \dots & 0 \\
        0 & 0 & \dots & 0 & \Psi_1 (x) & \Psi_2 (x) & \dots & \Psi_n (x)
    \end{bmatrix} 
\end{equation*}
contains the $n$ chosen basis functions repeated in each orthogonal direction; and $\mathbf{q}^{\epsilon i}$ is a column vector containing the $2n$ elastic coordinates of $B_i$. The basis functions must be twice differentiable and satisfy the boundary conditions $\frac{\p \Psi_i(x)}{\p x} \rvert_{x=0} = 0$ and $\frac{\p^2 \Psi_i(x)}{\p x^2}\rvert_{x=0} = 0$ to match the root condition of a cantilevered beam. The three following basis functions are chosen for this preliminary work, $\Psi_1(x)=x^2$, $\Psi_1(x)=x^3$, and $\Psi_1(x)=x^4$. Any number of functions obeying the aforementioned boundary conditions may be selected, so $\mbs{\Psi}(x)$ will be used to preserve generality.

\par The generalized coordinates for the unconstrained boom component $B_{i}$ are $\mathbf{q}^{B_i^\trans} = \Big[
        \mathbf{r}_a^{ia^\trans} \,\,\, \mathbf{p}^{ia^\trans} \,\,\, \mathbf{q}^{\epsilon i^\trans}
    \Big]$, 
where $\mathbf{p}^{ia}$ is the vectorized DCM $\mbf{C}_{ia}$, $\mathbf{r}_a^{ia}$ is the position of the base of the boom with respect to an unforced particle $a$ resolved in inertial frame $\mathcal{F}_{a}$, and  $\mathbf{q}^{\epsilon i}$ is the elastic coordinates of $B_i$. The generalized velocities for the unconstrained boom component $B_{i}, \text{ where } i \in \left\{ c, d, e, f\right\}$ are $\mbs{\nu}^{B_i^\trans} =
\Big[
    \dot{\mbf{r}}_a^{ia^\trans} \,\,\,
    \mbs{\omega}_i^{ia^\trans} \,\,\,
    \dot{\mbf{q}}^{\epsilon i^\trans}
\Big]$, 
where $\mbs{\omega}_i^{ia}$ is the angular velocity vector of boom body frame $\mathcal{F}_{i}$ relative to  $\mathcal{F}_{a}$, resolved in $\mathcal{F}_{i}$. These generalized velocities are related to the generalized coordinates through the mappings
\begin{align}
    \dot{\mbf{q}}^{B_i} &= \mbs{\Gamma}_{B_i}\mbs{\nu}^{B_i}, \label{eq:BoomKinMap1}\\
    \mbs{\nu}^{B_i} &= \mbf{S}_{B_i} \dot{\mbf{q}}^{B_i},\label{eq:BoomKinMap2}
\end{align}
where $\mbs{\Gamma}_{B_i} = \text{diag}\{ \mbf{1}, \;  \mbs{\Gamma}^{ia}_i(\mbf{p}^{ia}), \; \mbf{1} \}$, $\mbf{S}_{B_i} = \text{diag}\{ \mbf{1}, \;  \mbf{S}^{ia}_i(\mbf{p}^{ia}), \; \mbf{1} \}$, and $\mbf{S}_{B_i}\mbs{\Gamma}_{B_i} = \mbf{1}$.

\par The deformed position of a mass element $\mathrm{d}m$ at length $x$ along boom $B_i$ relative to point $a$ and resolved in $\mathcal{F}_a$ is given by
\bdis
    \mbf{r}_a^{\mathrm{d}ma} = \mbf{r}_a^{ia} + \mbf{C}_{ia}^\trans \mbf{r}_i^{\mathrm{d}mi},
\edis
where $\mbf{r}_i^{\mathrm{d}mi^\trans} = \Big[ x \,\,\, \mbf{u}^{i^\trans}(x,t) \Big] = \Big[ x \,\,\, \left(\mbs{\Psi}(x)\mbf{q}^{\epsilon i}(t)\right)^\trans \Big]$.

\section{Component Kinetic \& Potential Energies}
\par Kinetic energy is modeled for all components with the respective generalized velocities. It is assumed that all components exist outside of any gravity field, so potential energy is only modeled for the CABLESSail boom components in the form of strain energy.
\subsection{Solar Sail Bus}
\par The kinetic energy of the Solar sail bus, body $B_b$, is
\begin{equation}
\label{eq:KE_Hub}
    T_{B_b a/a} = \frac{1}{2} m_{B_b}\dot{\mathbf{r}}_a^{ba^\trans} \dot{\mathbf{r}}_a^{ba} + \frac{1}{2} \mathbf{\omega}_b^{ba^\trans} \mathbf{I}_b^{B_b b} \mathbf{\omega}_b^{ba} = 
    \frac{1}{2}
    \mbs{\nu}^{B_b^\trans}
    \mbf{M}_{B_b}
    \mbs{\nu}^{B_b} = \frac{1}{2}\dot{\mbf{q}}^{B_b^\trans}\mbf{S}_{B_b}^\trans\mbf{M}_{B_b}\mbf{S}_{B_b}\dot{\mbf{q}}^{B_b},
\end{equation}
where $m_{B_b}$ is the mass of body $B_b$, $\mathbf{I}_b^{B_b b} $ is the moment of inertia matrix of body $B_b$ relative to point $b$ resolved in $\mathcal{F}_{b}$, and the mass matrix is given by $\mbf{M}_{B_b}= \text{diag}\{
        m_{B_b}\mbf{1},\mathbf{I}_b^{B_b b}\}$. 

\subsection{Spacecraft Bus}
\label{sec: spacecraft bus energies}
The kinetic energy of the spacecraft bus, body $B_g$, are derived similarly to the solar sail bus, resulting in
\begin{equation}
\label{eq:KE_AMT}
    T_{B_g a/a} = \frac{1}{2} m_{B_g}\dot{\mathbf{r}}_a^{ga^\trans} \dot{\mathbf{r}}_a^{ga} + \frac{1}{2} \mathbf{\omega}_g^{ga^\trans} \mathbf{I}_g^{B_g g} \mathbf{\omega}_g^{ga} = 
    \frac{1}{2}
    \mbs{\nu}^{B_g^\trans}
    \mbf{M}_{B_g}
    \mbs{\nu}^{B_g} = \frac{1}{2}\dot{\mbf{q}}^{B_g^\trans}\mbf{S}_{B_g}^\trans\mbf{M}_{B_g}\mbf{S}_{B_g}\dot{\mbf{q}}^{B_g},
\end{equation}
where $m_{B_g}$ is the mass of body $B_g$, $\mathbf{I}_b^{B_g g} $ is the moment of inertia matrix of body $B_g$ relative to point $g$ resolved in $\mathcal{F}_{g}$, and the mass matrix is given by $\mbf{M}_{B_g}= \text{diag}\{
        m_{B_g}\mbf{1},\mathbf{I}_g^{B_g g}\}$.

\subsection{CABLESSail Booms}
\par The flexible boom is modeled as an Euler-Bernoulli beam and therefore has a strain energy due to bending in its two transverse directions given by~\cite{Rao2019}
\begin{equation}
    V_{B_i} = \frac{1}{2}EI \int_{0}^{L} \left[\left( \frac{\partial^2 u_2(x)}{\partial x^2}\right)^2 + \left(\frac{\partial^2 u_3(x)}{\partial x^2}\right)^2\right] \mathrm{d}x
    ,
\end{equation}
where $E$ is the modulus of elasticity of the boom and $I$ is the second moment of area of the boom cross section.
Using the deflection definition in \eqref{eq: deflection definition}, the strain energy of body $B_i$
\begin{equation}
\label{eq:PE_Boom}
    V_{B_i} = \onehalf \mathbf{q}^{\epsilon i^\trans }\left(EI \int^L_0 \left( \frac{\partial^2\mathbf{\Psi}}{\partial x^2}    \right)^\trans
    \left( \frac{\partial^2\mathbf{\Psi}}{\partial x^2}    \right)\mathrm{d}x\right)
    \mathbf{q}^{\epsilon i} = \frac{1}{2} \mbf{q}^{B_i^\trans} \mbf{K}_{B_i} \mbf{q}^{B_i},
\end{equation}
where $\mbf{K}_{B_i} = \text{diag}\left\{ \mbf{0}, \quad \mbf{0} , \quad EI \int^L_0 \left( \frac{\partial^2\mathbf{\Psi}}{\partial x^2}    \right)^\trans
    \left( \frac{\partial^2\mathbf{\Psi}}{\partial x^2}    \right)\mathrm{d}x  \right\}$.

\par The kinetic energy of the boom $B_i$ is the integral along the boom of the kinetic energy of every mass element, $\mathrm{d}m$, and is computed as 
\begin{equation}\label{eq:boomKE}
    T_{B_ia/a} = \frac{1}{2}\int_\mathcal{B} \vect{r}^{\mathrm{d}ma^{\cdot a}}\cdot\vect{r}^{\mathrm{d}ma^{\cdot a}} \mathrm{d}m =  \frac{1}{2}\int_\mathcal{B} \vect{v}^{\mathrm{d}ma/a} \cdot \vect{v}^{\mathrm{d}ma/a} \mathrm{d}m ,
\end{equation}
where $\vect{r}^{\mathrm{d}ma} = \vect{r}^{\mathrm{d}mi} + \vect{r}^{ia}$. Note that $\vect{r}^{ia}$ is the location of the base of the boom $B_i$ relative to point $a$ and $\vect{r}^{\mathrm{d}mi}$ is the location of the mass element relative to the base of the boom. The deflections are used to define the mass element position, where $\mathbf{r}^{\mathrm{d}mi^\trans}_i = \Big[
        x \,\,\,
        \left(\mathbf{\Psi}(x) \mathbf{q}^{\epsilon i} \right)^\trans
    \Big]$.
After taking the time derivative of $\vect{r}^{\mathrm{d}ma}$ with respect to $\mathcal{F}_a$, the kinetic energy of the boom can be factored into the quadratic form
\begin{equation}
\label{eq:KE_Boom}
    T_{B_iw/a} = \frac{1}{2}
    \mbs{\nu}^{B_i^\trans}
    \mathbf{M}_{B_i}
    \mbs{\nu}^{B_i},
\end{equation}
where the mass matrix $\mathbf{M}_{B_i}$ is given by
\begin{equation}\label{M_matrix_booms}
    \mathbf{M}_{B_i} = 
    \begin{bmatrix}
        m_{B_i}\mathbf{1} && -\mathbf{C}_{ia}^\trans \mathbf{c}_i^{B_ii^\times} && \mathbf{C}_{ia}^\trans \mbf{H}_{B_i} \\
        -\mathbf{c}_i^{{B_ii^\times}^\trans}\mathbf{C}_{ia}  &&  \mbf{I}^{B_ii}_i  &&  -\mbf{G}_{B_i} \\
        \mbf{H}_{B_i}^\trans  \mathbf{C}_{ia} &&  -\mbf{G}_{B_i}^\trans   &&  \mbf{P}_{B_i}
    \end{bmatrix}.
\end{equation}
The contents of the mass matrix include the boom's total mass, $m_{B_i}$, the boom's first moment of mass relative to point $i$ resolved in $\mathcal{F}_i$, $\mathbf{c}_i^{B_ii}$, the boom's second moment of mass relative to point $i$ resolved in $\mathcal{F}_i$, $\mbf{I}^{B_ii}_i$, and other matrices associated with the elastic coordinates, $\mbf{H}_{B_i}$, $\mbf{G}_{B_i}$, and $\mbf{P}_{B_i}$.  Detailed derivations of these terms are found in Appendix~A of~\cite{bunker2024modular}, where it is also shown that the time derivative of $\mathbf{M}_{B_i}$ is
\begin{equation}\label{M_Bi_dot_booms}
    \dot{\mathbf{M}}_{B_i} = 
    \begin{bmatrix}
        \mathbf{0} && (\mbs{\omega}^{ia^\times}_i\mbf{C}_{ia})^\trans \mathbf{c}_i^{B_ii^\times} && -(\mbs{\omega}^{ia^\times}_i\mbf{C}_{ia})^\trans \mbf{H}_{B_i} \\
        \mathbf{c}_i^{{B_ii^\times}^\trans} \mbs{\omega}^{ia^\times}_i\mbf{C}_{ia} &&  \mathbf{0} &&  \mathbf{0}  \\
        -\mbf{H}_{B_i}^\trans \mbs{\omega}^{ia^\times}_i\mbf{C}_{ia} &&  \mathbf{0} &&  \mathbf{0} 
    \end{bmatrix}.
\end{equation}

\par To assist with the derivation of the boom's equations of motion, it is convenient to substitute the kinematic mapping~\eqref{eq:BoomKinMap2} into~\eqref{eq:KE_Boom}, resulting in the kinetic energy expression
\begin{equation}
\label{eq:KE_Boom2}
    T_{B_ia/a} = \frac{1}{2}\dot{\mbf{q}}^{B_i^\trans}\mbf{S}_{B_i}^\trans\mbf{M}_{B_i}\mbf{S}_{B_i}\dot{\mbf{q}}^{B_i}.
\end{equation}

\section{Component Generalized Forces}
\par External forces that do work on the system are considered with the null-space method. Thus, the SRP force and torque are considered for the solar sail bus, and the reaction forces due to cable tensioning are considered for the booms. No other forces are considered for this work, but the methods presented here may easily be adapted for other applications.

\subsection{Solar Sail Bus}

\par The solar sail membrane detailed in Section \ref{sec:SolarSailModeling} produces an external force and torque pair that is applied at the center of mass of the solar sail bus, which is assumed to be coincident with the center of the solar sail membrane if it were perfectly flat with undeformed booms. To preserve generality, a generic external force is considered.

\par The virtual work done by the external force $\vect{f}^{B_b}$ acting on the center of mass of the solar sail bus is
\begin{equation}
\label{eq:GenForcesExt1a}
    \delta W^{B_b f} = \vect{f}^{B_b} \cdot \vect{r}^{ba} = \mbf{f}_a^{B_b^\trans} \delta \mbf{r}_a^{ba} = \mbf{f}_b^{B_b^\trans} \mbf{C}_{ba} \delta \mbf{r}_a^{ba} = \delta \mbf{q}^{B_i^\trans} \bbm \mbf{C}_{ba}^\trans\mbf{f}_b^{B_b} \\ \mbf{0} \ebm,
\end{equation}
where $\mbf{f}_b^{B_b}$ contains the components of $\vect{f}^{B_b}$ resolved in $\mathcal{F}_b$.

The virtual work done by the external torque $\vect{\tau}^{B_b}$ acting on the solar sail bus is
\begin{equation}
\label{eq:GenForcesExt1}
    \delta W^{B_b \tau} = 
    \mbs{\tau}_b^{B_b^\trans} \delta \mbs{\theta},
\end{equation}
where
\begin{equation}
\label{eq:GenForcesExt2}
    \delta \mbs{\theta} = \frac{\p \mbs{\omega}^{ba}_b }{\p \dot{\mbf{p}}^{ba} } \delta \mbf{p}^{ba} 
    =  \frac{\p  }{\p \dot{\mbf{p}}^{ba} } 
   \left(\mbf{S}(\mbf{p}^{ba}) \dot{\mbf{p}}^{ba}  \right)
   \delta \mbf{p}^{ba}
   = \mbf{S}(\mbf{p}^{ba}) \delta \mbf{p}^{ba}.
\end{equation}
Substituting~\eqref{eq:GenForcesExt2} into~\eqref{eq:GenForcesExt1} results in
\begin{equation}
\label{eq:GenForcesExt3}
    \delta W^{B_b \tau} = \mbs{\tau}_b^{B_b^\trans} \mbf{S}(\mbf{p}^{ba}) \delta \mbf{p}^{ba} = \delta \mbf{q}^{B_i^\trans} \bbm \mbf{0} \\ \mbf{S}^\trans(\mbf{p}^{ba})\mbs{\tau}_b^{B_b}  \ebm.
\end{equation}
Combining the results from~\eqref{eq:GenForcesExt1a} and~\eqref{eq:GenForcesExt3}, the generalized forces acting on $B_b$ due to the external force and torque are
\begin{equation*}
    \mbf{f}_{B_i} = \bbm \mbf{C}_{ba}^\trans\mbf{f}_b^{B_b} \\ \mbf{S}^\trans(\mbf{p}^{ba})\mbs{\tau}_b^{B_b}  \ebm.
\end{equation*}

\subsection{CABLESSail Booms}

Each of the booms, denoted by bodies $B_i$, $i\in \{c,d,e,f\}$, have actuating cables routed through them, as shown in Fig.~\ref{fig:CABLESSail boom diagram}. When tension is applied to the cables, they impart reaction forces on the boom, which then results in bending of the boom. To provide full actuation capabilities in each of the bending directions, four cables are included: one along the ``top'' of the boom (nominally a distance $R+h$ in the positive $\vect{i}_3$ from the boom's neutral axis), one along the ``bottom'' of the boom (nominally a distance $R+h$ in the negative $\vect{i}_3$ from the boom's neutral axis), as well as cables along the ``sides'' of the boom (nominally distances $R+h$ in the positive and negative $\vect{i}_2$ from the boom's neutral axis). The $j^\mathrm{th}$ cable, $j \in \{1,2,3,4\}$ passes through routing holes located at points $l_{j,k}$, $k=1,\ldots,n-1$ and is attached to the end of the boom at the point $l_{j,n}$. The position of points $l_{j,k}$ relative to point $i$ at the root of the boom is given by
\beq
\label{eq:CablePos1}
\mbf{r}_i^{l_{j,k}i} = \mbf{r}_\mbf{r}^{p_{k}i} + \mbf{C}_{p_ki}^\trans \mbf{r}_{p_k}^{l_{j,k}p_k},
\eeq
where $\mbf{r}_\mbf{r}^{p_{k}i^\trans} = \Big[ x_k \,\,\, \mbs{\Psi}(x_k) \mbf{q}^{\epsilon i} \Big]$ is the deformed position of the beam at the spreader routing hole a distance $x_k$ down the length of the boom, $\mbf{C}_{p_ki} \approx \mbf{1} - \mbs{\theta}^{p_ki^\times}$ is the small angle approximation of the deformed attitude of the boom at a length $x_k$ down its length with $\mbs{\theta}^{p_ki} = \Big[ 0 \,\,\, -\left(\frac{\p \mbs{\Psi}_{u_3}}{\p x} \mbf{q}^{\epsilon i}\right)^\trans \,\,\, \left(\frac{\p \mbs{\Psi}_{u_2}}{\p x}  \mbf{q}^{\epsilon i}\right)^\trans \Big]$~\cite{Hughes2004,damaren1995simulation}, and $\mbf{r}_{p_k}^{l_{j,k}p_k}$ is the position of the cable routing point relative to the deformed point $p_k$ with
\bdis
\mbf{r}_{p_k}^{l_{1,k}p_k^\trans} = \Big[ 0 \,\,\, 0 \,\,\, R+h \Big], \,\, \mbf{r}_{p_k}^{l_{2,k}p_k^\trans} = \Big[ 0 \,\,\, 0 \,\,\, -(R+h) \Big], \,\, \mbf{r}_{p_k}^{l_{3,k}p_k^\trans} = \Big[ 0 \,\,\, R+h \,\,\, 0  \Big], \,\, \mbf{r}_{p_k}^{l_{4,k}p_k^\trans} = \Big[ 0 \,\,\, -(R+h) \,\,\, 0 \Big].
\edis
In this note, it is assumed that the spreaders are equally spaced along the length of the boom, such that $\Delta x = x_{k+1} - x_{k}$, $k=1,\ldots,n-1$ is a constant distance.

It is assumed that there is no friction between the cable surface and the routing hole in the spreader plate, so the only reaction forces are perpendicular to the deflected boom. The reaction force for the $j^\mathrm{th}$ cable at location $l_{j,k}$ is
\begin{equation*}
    \mbf{f}_{i}^{B_if_{j,k}} =\frac{T_j}{\Delta x} \left(\mbf{r}_i^{l_{j,k+1}i} - 2\mbf{r}_i^{l_{j,k}i} + \mbf{r}_i^{l_{j,k-1}i}\right), \quad j=1,2,3,4, \quad k=1,\ldots,n-1,
\end{equation*}
where $T_j$ is the tension within the $j^\mathrm{th}$ cable and $x_0 = 0$. The reaction force due to the $j^\mathrm{th}$ cable at the end of the boom is given by
\begin{equation*}
    \mbf{f}_{i}^{B_if_{j,n}} =\frac{T_j}{\Delta x} \left(\mbf{r}_i^{l_{j,n}i} - \mbf{r}_i^{l_{j,n-1}i} \right), \quad j=1,2,3,4.
\end{equation*}

The cable reaction force $\vect{f}^{B_if_{j,k}}$ is applied at point $l_{j,k}$ whose position relative to point $a$ resolved in $\mathcal{F}_a$ is
\beq
\label{eq:PosCableAttach}
\mbf{r}_a^{l_{j,k}a} = \mbf{r}^{ia}_a + \mbf{C}_{ia}^\trans \mbf{r}_\mbf{r}^{p_{k}i} + \mbf{C}_{ia}^\trans\mbf{C}_{p_ki}^\trans \mbf{r}_{p_k}^{l_{j,k}p_k}.
\eeq

The virtual work done by $\vect{f}^{B_if_{j,k}}$ is
\begin{equation}\label{EQ:VirtualWork}
    \delta W^{B_if_{j,k}}
    =\vect{f}^{B_if_{j,k}}\cdot \delta \vect{r}^{l_{jk}a} = \mbf{f}^{B_if_{j,k}^\trans}_a \delta \mbf{r}^{l_{j,k}a}_a = \mbf{f}^{B_if_{j,k}^\trans}_{i} \mbf{C}_{ia}\delta \mbf{r}^{l_{j,k}a}_a.
\end{equation}
The computation of the virtual displacement $\delta \mbf{r}^{l_{j,k}a}_a$ is described in Appendix~B of~\cite{bunker2024modular}, where it is shown that~\eqref{EQ:VirtualWork} is related to a virtual displacement in the generalized coordinates through the equation 
\begin{equation*}
    \delta W^{B_if_{j,k}} = \delta \mbf{q}^{B_i^\trans} \bbm \mbf{C}_{ia}^\trans \mbf{f}^{B_if_{j,k}}_{i} \\ \left(\frac{\p}{\p \mbf{p}^{ia}}\left(\mbf{C}_{ia}^\trans \left(\mbf{r}^{p_ki}_i + \mbf{C}_{p_ki}^\trans\mbf{r}^{l_{j,k}p_k}_l\right)\right)\right)^\trans \mbf{C}_{ia}^\trans \mbf{f}^{B_if_{j,k}}_{i} \\ \left(\bbm \mbf{0} \\ \mbs{\Psi}(x_k) \ebm^\trans +\bbm \mbf{0} \\ -\frac{\p \mbs{\Psi}_{u_3}}{\p x}  \\ \frac{\p \mbs{\Psi}_{u_2}}{\p x}   \ebm^\trans\mbf{r}_{p_k}^{l_{j,k}p_k^\times} \right) \mbf{f}^{B_if_{j,k}}_{i} \ebm.
\end{equation*}
Therefore the generalized forces due to the cable reaction forces on body $B_i$ are
\bdis
\mbf{f}_{B_i} = \sum_{j=1}^4\sum_{k=0}^n\bbm \mbf{C}_{ia}^\trans\mbf{f}^{B_if_{j,k}}_{i} \\ \left(\frac{\p}{\p \mbf{p}^{ia}}\left(\mbf{C}_{ia}^\trans \left(\mbf{r}^{p_ki}_i + \mbf{C}_{l_ki}^\trans\mbf{r}^{l_kp_k}_l\right)\right)\right)^\trans \mbf{C}_{ia}^\trans \mbf{f}^{B_if_{j,k}}_{i} \\ \left(\bbm \mbf{0} \\ \mbs{\Psi}(x_k) \ebm^\trans +\bbm \mbf{0} \\ -\frac{\p \mbs{\Psi}_{u_3}}{\p x}  \\ \frac{\p \mbs{\Psi}_{u_2}}{\p x}   \ebm^\trans\mbf{r}_l^{l_kp_k^\times} \right) \mbf{f}^{B_if_{j,k}}_{i} \ebm.
\edis

\section{Lagrange's Equations for Each Component}

Lagrange's equation is used to derive the equations of motion for each component. The general form of Lagrange's equation is given as
\begin{equation}\label{eq: Lagranges equation}
    \frac{\mathrm{d}}{\mathrm{d}t} \left(  \frac{\p {L}}{\p \dot{\mbf{q}} } \right)^\trans 
    - \left( \frac{\p {L}}{\p \mbf{q}}\right)^\trans 
    = \mbf{f} + \mbs{\Xi}^\trans \mbs{\lambda}
    ,
\end{equation}
where $L = T - V$ is the Lagrangian, $T$ is the kinetic energy, $V$ is the potential energy, $\mbf{q}$ contains the generalized coordinates, $\mbf{f}$ is the generalized forces, $\mbs{\Xi}$ is the constraint matrix associated with the constraints $\mbs{\Xi}\dot{\mbf{q}} = \mbf{0}$, and $\mbs{\lambda}$ are Lagrange multipliers.

\subsection{CABLESSail Booms}
For boom body $B_i$, the Lagrangian is $L_{B_ia/a} = T_{B_i a/a} - V_{B_i}$, $\mbf{q}^{B_i}$ are the generalized coordinates, $\mbf{f}_{B_i}$ are the generalized forces, $\mbs{\Xi}_{B_i}$ is the constraint matrix associated with the constrained attitude parameters, and $\mbs{\lambda}_{B_i}$ are Lagrange multipliers. Since the potential energy does not depend on the generalized velocities, Lagrange's equation for body $B_i$  can then be written as
\begin{equation}\label{eq: Expanded Lagranges equation}
    \frac{\mathrm{d}}{\mathrm{d}t} \left(  \frac{\p T_{B_i a/a}}{\p \dot{\mbf{q}}^{B_i} } \right)^\trans 
    - \left( \frac{\p T_{B_i a/a}}{\p \mbf{q}^{B_i}}\right)^\trans 
    + \left( \frac{\p V_{B_i} }{\p \mbf{q}^{B_i}}\right)^\trans 
    = \mbf{f}_{B_i} + \mbs{\Xi}_{B_i} \mbs{\lambda}_{B_i}.
\end{equation}

\par The partial derivatives of the boom's kinetic and potential energy are solved using the expressions in~\eqref{eq:PE_Boom} and~\eqref{eq:KE_Boom2}, resulting in
\begin{multline}
    \frac{\mathrm{d}}{\mathrm{d}t} \left(  \frac{\p T_{B_i a/a}}{\p \dot{\mbf{q}}^{B_i} } \right)^\trans 
    - \left( \frac{\p T_{B_i a/a}}{\p \mbf{q}^{B_i}}\right)^\trans 
    + \left( \frac{\p V_{B_i} }{\p \mbf{q}^{B_i}}\right)^\trans \\ = \mbf{S}^\trans_{B_i} \mbf{M}_{B_i} \dot{\mbs{\nu}}^{B_i} +
    \mbf{S}_{B_i}^\trans \dot{\mbf{M}}_{B_i} \mbs{\nu}^{B_i}
    +
    \left(\dot{\mbf{S}}_{B_i}^\trans - \mbshat{\Omega}_{B_i}\right) \mbf{M}_{B_i} \mbs{\nu}^{B_i}  
    - \bbm
        \mbf{0}  \\   
        \left(\frac{\hat{\p} \left( \mbf{C}_{ia}^\trans\mbf{v}^{B_i} \right)}{\hat{\p} \mbf{p}^{ia}}\right)^\trans \dot{\mbf{r}}_a^{ia} \\
        \mbf{0}
    \ebm
    +\mbf{K}_{B_i} \mbf{q}^{B_i},
    \label{unconstrained_component_eom}
\end{multline}
where $\mbshat{\Omega}_{B_i} = \text{diag}\left\{ \mbf{0} , \left( \frac{\p \mbs{\omega} ^{ia}_{i} }{\p \mbf{p}^{ia} } \right)^\trans ,  \mbf{0} \right\}$, $\mbf{v}^{B_i} =  \mbf{H}_{B_i}\dot{\mbf{q}}^{\epsilon i} -\mbf{c}_i^{B_ii^\times}\mbs{\omega}_i^{ia}$, and $\frac{\hat{\p} \left( \mbf{C}_{ia}^\trans\mbf{v}^{B_i} \right)}{\hat{\p} \mbf{p}^{ia}}$ is the partial derivative of $\mbf{C}_{ia}^\trans\mbf{v}^{B_i}$ with respect to $\mbf{p}^{ia}$ considering $\mbf{v}^{B_i}$ as a constant. 
Details regarding the computation of these partial derivatives are provided in Appendix~C of~\cite{bunker2024modular}.

\subsection{Solar Sail Bus}

The Lagrangian of the rigid hub $B_b$ is $L_{B_ba/a} = T_{B_ba/a}$, its generalized coordinates are $\mbf{q}^{B_b}$, resulting in Lagrange's equation for body $B_b$ in the form
\begin{equation}
\label{eq: Expanded Lagranges equation2}
    \frac{\mathrm{d}}{\mathrm{d}t} \left(  \frac{\p T_{B_b a/a}}{\p \dot{\mbf{q}}^{B_b} } \right)^\trans 
    - \left( \frac{\p T_{B_b a/a}}{\p \mbf{q}^{B_b}}\right)^\trans 
    = \mbf{f}_{B_b} + \mbs{\Xi}_{B_b} \mbs{\lambda}_{B_b}.
\end{equation}


A similar procedure to that used for the CABLESSail booms is performed on the spacecraft hub's kinetic energy in~\eqref{eq:KE_Hub}, which results in
\begin{equation} \label{eq:unconstrained_component_eom2}
    \frac{\mathrm{d}}{\mathrm{d}t} \left(  \frac{\p T_{B_b a/a}}{\p \dot{\mbf{q}}^{B_b} } \right)^\trans 
    - \left( \frac{\p T_{B_b a/a}}{\p \mbf{q}^{B_b}}\right)^\trans 
     = \mbf{S}^\trans_{B_b} \mbf{M}_{B_b} \dot{\mbs{\nu}}^{B_b} +
    \mbf{S}_{B_b}^\trans \dot{\mbf{M}}_{B_b} \mbs{\nu}^{B_b} +
    \left(\dot{\mbf{S}}_{B_b}^\trans - \mbshat{\Omega}_{B_b}\right)\mbf{M}_{B_b} \mbs{\nu}^{B_b},
\end{equation}
where $\mbshat{\Omega}_{B_b} = \text{diag}\left\{ \mbf{0} , \left( \frac{\p \mbs{\omega} ^{ba}_{b} }{\p \mbf{p}^{ba} } \right)^\trans\right\}$. 
Details of this computation are also provided in Appendix~C of~\cite{bunker2024modular}.

\subsection{Spacecraft Bus}
The equations of motion of both buses are identical up to specific dimension and inertia values. Therefore the resulting Lagrange's equation of motion for the spacecraft bus, $\mathcal{B}_?$, are nearly identical to \eqref{eq:unconstrained_component_eom2}, resulting in 
\begin{equation} \label{eq:unconstrained_component_eom3}
    \frac{\mathrm{d}}{\mathrm{d}t} \left(  \frac{\p T_{B_g a/a}}{\p \dot{\mbf{q}}^{B_g} } \right)^\trans 
    - \left( \frac{\p T_{B_g a/a}}{\p \mbf{q}^{B_g}}\right)^\trans 
     = \mbf{S}^\trans_{B_g} \mbf{M}_{B_g} \dot{\mbs{\nu}}^{B_g} +
    \mbf{S}_{B_g}^\trans \dot{\mbf{M}}_{B_g} \mbs{\nu}^{B_g} +
    \left(\dot{\mbf{S}}_{B_g}^\trans - \mbshat{\Omega}_{B_g}\right)\mbf{M}_{B_g} \mbs{\nu}^{B_?},
\end{equation}

\section{Attitude Parameterization Constraint} \label{sec:Attitude Parameterization Constraint}
To ensure a valid attitude parameterization, the vectorized DCM associated with each of the components must be constrained through the Lagrange multipliers $\lambda_i$, $i\in \{b,c,d,e,f,g\}$. As shown in \cite{de_ruiter_SO3parameterixation_identities_2014}, the orthonormal property of the DCM results in its vectorized parameters satisfying the constraints
\begin{equation*}
    \mbs{\Phi}(\mathbf{p}^{ia}) = \begin{bmatrix}
        \mbf{p}_1^{ia^\trans}\mbf{p}_1^{ia} - 1\\
        \mbf{p}_2^{ia^\trans}\mbf{p}_2^{ia} - 1\\
        \mbf{p}_1^{ia^\trans}\mbf{p}_2^{ia} \\
        \mbf{p}_1^{ia^\times}\mbf{p}_2^{ia} - \mbf{p}_3^{ia}
    \end{bmatrix}
    = \mbs{0},
\end{equation*}
which can alternatively be written at the rate level as
\beq
\label{eq:RateConstraint}
\mbs{\Xi}^{ia}_i\dot{\mbf{p}}^{ia}  =\mbf{0},
\eeq
where
\begin{equation}
\label{eq:AttContraintMatrix}
    \mbs{\Xi}^{ia}_i  = 
    \begin{bmatrix}
        \mbf{p}_1^{ia^\trans}   && \mbf{0}                  && \mbf{0}  \\ 
        \mbf{0}                 && \mbf{p}_2^{ia^\trans}    && \mbf{0} \\
        \mbf{p}_2^{ia^\trans}   && \mbf{p}_1^{ia^\trans}    && \mbf{0} \\
        -\mbf{p}_2^{ia^\times}  && \mbf{p}_1^{ia^\times}    && -\mbf{1}
    \end{bmatrix}.
\end{equation}
The constraint matrix in~\eqref{eq:AttContraintMatrix} is related to the constraint matrices in~\eqref{eq: Expanded Lagranges equation} and~\eqref{eq: Expanded Lagranges equation2} through the relationships
\begin{align}
    \mbs{\Xi}_{B_i} &= \Big[ \mbf{0} \,\,\, \mbs{\Xi}^{ia}_i \,\,\, \mbf{0} \Big], \quad i\in \{c,d,e,f\}, \\
    \mbs{\Xi}_{B_b} &= \Big[ \mbf{0} \,\,\, \mbs{\Xi}^{ba}_b \Big],
\end{align}
which allows for~\eqref{eq:RateConstraint} to be rewritten as
\begin{align}
    \mbs{\Xi}_{B_i} \dot{\mbf{q}}^{B_i}&= \mbf{0}, \quad i\in \{c,d,e,f\}, \label{eq:RateConstraint1}\\
    \mbs{\Xi}_{B_b} \dot{\mbf{q}}^{B_b}&= \mbf{0}. \label{eq:RateConstraint2}
\end{align}
Substituting~\eqref{eq:HubKinMap1} and~\eqref{eq:BoomKinMap1} into~\eqref{eq:RateConstraint1} and~\eqref{eq:RateConstraint2}, respectively, results in the properties $\mbs{\Xi}_{B_i} \mbs{\Gamma}_{B_i}= \mbf{0}$, $i \in \{c,d,e,f\}$ and $\mbs{\Xi}_{B_b} \mbs{\Gamma}_{B_b}= \mbf{0}$.  This demonstrates how $\mbs{\Gamma}_{B_i}$ and $\mbs{\Gamma}_{B_b}$ lie in the nullspace of the constraint matrices, which forms the basis of the null-space constraint technique.

The method proceeds for the boom equations of motion by substituting~\eqref{unconstrained_component_eom} into~\eqref{eq: Expanded Lagranges equation}, then pre-multiplying on the left by $\mbs{\Gamma}^{\trans}_{B_i}$, resulting in
\begin{multline}
    \label{eq:ComponentEOMNullSpace1}
    \mbs{\Gamma}^{\trans}_{B_i}\mbf{S}^\trans_{B_i} \mbf{M}_{B_i} \dot{\mbs{\nu}}^{B_i} +
    \mbs{\Gamma}^{\trans}_{B_i}\mbf{S}_{B_i}^\trans \dot{\mbf{M}}_{B_i} \mbs{\nu}^{B_i} +
    \mbs{\Gamma}^{\trans}_{B_i}\left(\dot{\mbf{S}}_{B_i}^\trans - \mbshat{\Omega}_{B_i}\right)\mbf{M}_{B_i} \mbs{\nu}^{B_i} 
    - \mbs{\Gamma}^{\trans}_{B_i}\bbm
        \mbf{0}  \\   
        \left(\frac{\hat{\p} \left( \mbf{C}_{ia}^\trans\mbf{v}^{B_i} \right)}{\hat{\p} \mbf{p}^{ia}}\right)^\trans \dot{\mbf{r}}_a^{ia} \\
        \mbf{0}
    \ebm
    +\mbs{\Gamma}^{\trans}_{B_i}\mbf{K}_{B_i} \mbf{q}^{B_i}
    \\ = \mbs{\Gamma}^{\trans}_{B_i}\mbf{f}_{B_i} + \mbs{\Gamma}^{\trans}_{B_i}\mbs{\Xi}_{B_i}^\trans \mbs{\lambda}_{B_i}.
\end{multline}
We know that $ \mbs{\Gamma}_{B_i}\mbf{S}_{B_i} =\mbs{\Gamma}^{\trans}_{B_i}\mbf{S}^\trans_{B_i}  = \mbf{1}$ and $ \mbs{\Gamma}_{B_i}^\trans \mbs{\Xi}_{B_i}^\trans = \mbf{0} $, therefore,~\eqref{eq:ComponentEOMNullSpace1} is simplified to
\begin{equation}
\label{eq:ComponentEOMNullSpace2}
    \mbf{M}_{B_i} \dot{\mbs{\nu}}^{B_i}
    + \dot{\mbf{M}}_{B_i} \mbs{\nu}^{B_i} +\mbs{\Gamma}^{\trans}_{B_i}\left(\dot{\mbf{S}}_{B_i}^\trans - \mbshat{\Omega}_{B_i}\right)\mbf{M}_{B_i} \mbs{\nu}^{B_i}
    - \mbs{\Gamma}^{\trans}_{B_i}\bbm
        \mbf{0}  \\   
        \left(\frac{\hat{\p} \left( \mbf{C}_{ia}^\trans\mbf{v}^{B_i} \right)}{\hat{\p} \mbf{p}^{ia}}\right)^\trans \dot{\mbf{r}}_a^{ia} \\
        \mbf{0}
    \ebm
    + \mbs{\Gamma}^{\trans}_{B_i} \mbf{K}_{B_i} \mbf{q}^{B_i} 
    =
    \mbs{\Gamma}^{\trans}_{B_i}\mbf{f}_{B_i}.
\end{equation}
As a final step in the derivation of these equations of motion, the identities $\mbs{\Gamma}^{\trans}_{B_i}\left(\dot{\mbf{S}}_{B_i}^\trans - \mbshat{\Omega}_{B_i}\right) = \mbs{\Omega}_{B_i}$ and $\mbs{\Gamma}_i^{ia^\trans} \left(\frac{\hat{\p} \left( \mbf{C}_{ia}^\trans\mbf{v}^{B_i} \right)}{\hat{\p} \mbf{p}^{ia}}\right)^\trans  = \mbf{v}^{B_i^\times }\mbf{C}_{ia}$ from~\cite{de_ruiter_SO3parameterixation_identities_2014} are made use of, where $\mbs{\Omega}_{B_i} = \text{diag}\{\mbf{0},\mbs{\omega}_i^{ia^\times},\mbf{0}\}$. Applying these identities to~\eqref{eq:ComponentEOMNullSpace2} results in
\begin{equation}
\label{eq:ComponentEOMNullSpace3}
    \mbf{M}_{B_i} \dot{\mbs{\nu}}^{B_i}
     + \mbs{\Gamma}^{\trans}_{B_i} \mbf{K}_{B_i} \mbf{q}^{B_i} + \mbf{f}^{non}_{B_i}
    =
    \mbs{\Gamma}^{\trans}_{B_i}\mbf{f}_{B_i},
     \quad i \in \{d, e, f, g\},
\end{equation}
where $\mbf{f}^{non}_{B_i} = \left(\dot{\mbf{M}}_{B_i}  + \mbs{\Omega}_{B_i}\mbf{M}_{B_i}\right) \mbs{\nu}^{B_i} - \Big[
        \mbf{0}  \,\,\,   
        \left(\mbf{v}^{B_i^\times }\mbf{C}_{ia}\dot{\mbf{r}}_a^{ia}\right)^\trans \,\,\,
        \mbf{0}
    \Big]^\trans$.
The equations of motion of boom body $B_i$, $i\in \{c,d,e,f\}$ with the Lagrange multipliers removed are thus given by~\eqref{eq:ComponentEOMNullSpace3}. 

A similar process is used to obtain the equations of motion for both the solar sail and spacecraft buses, bodies $B_b, B_c$, in the form
\begin{equation}
\label{eq: Independent equations of motion busses}
    \mbf{M}_{B_i} \dot{\mbs{\nu}}^{B_i}
     + \mbf{f}^{non}_{B_i}
    =
    \mbs{\Gamma}^{\trans}_{B_i}\mbf{f}_{B_i}, \quad i \in \{b, g\},
\end{equation}
where $\mbf{f}^{non}_{B_b} = \left(\dot{\mbf{M}}_{B_b}  + \mbs{\Omega}_{B_b}\mbf{M}_{B_b}\right) \mbs{\nu}^{B_b}$ and $\mbs{\Omega}_{B_b} = \text{diag}\{\mbf{0},\mbs{\omega}_b^{ba^\times}\}$.

\section{Modular Multi-body Dynamic Modeling} 
\label{sec:Modular Multibody Dynamic Modeling}
\par With the dynamics of each individual component fully described in~\eqref{eq:ComponentEOMNullSpace3} and~\eqref{eq: Independent equations of motion busses} with the attitude parameterization Lagrange multipliers removed, the components can now be constrained together to ``assemble'' the full solar sail. First, all independent component equations of motion are aggregated into a single matrix expression. Then, the assembly constraints are defined and enforced, followed by the application of the null-space method to remove the corresponding Lagrange multipliers. This process is first shown in detail for a nominal assembly of CABLESSail, then an abbreviated description is given for a second assembly where the optional spacecraft bus is used to represent an active mass translator actuator.

\begin{figure}[t!]
    \centering

    \includegraphics[scale=0.35]{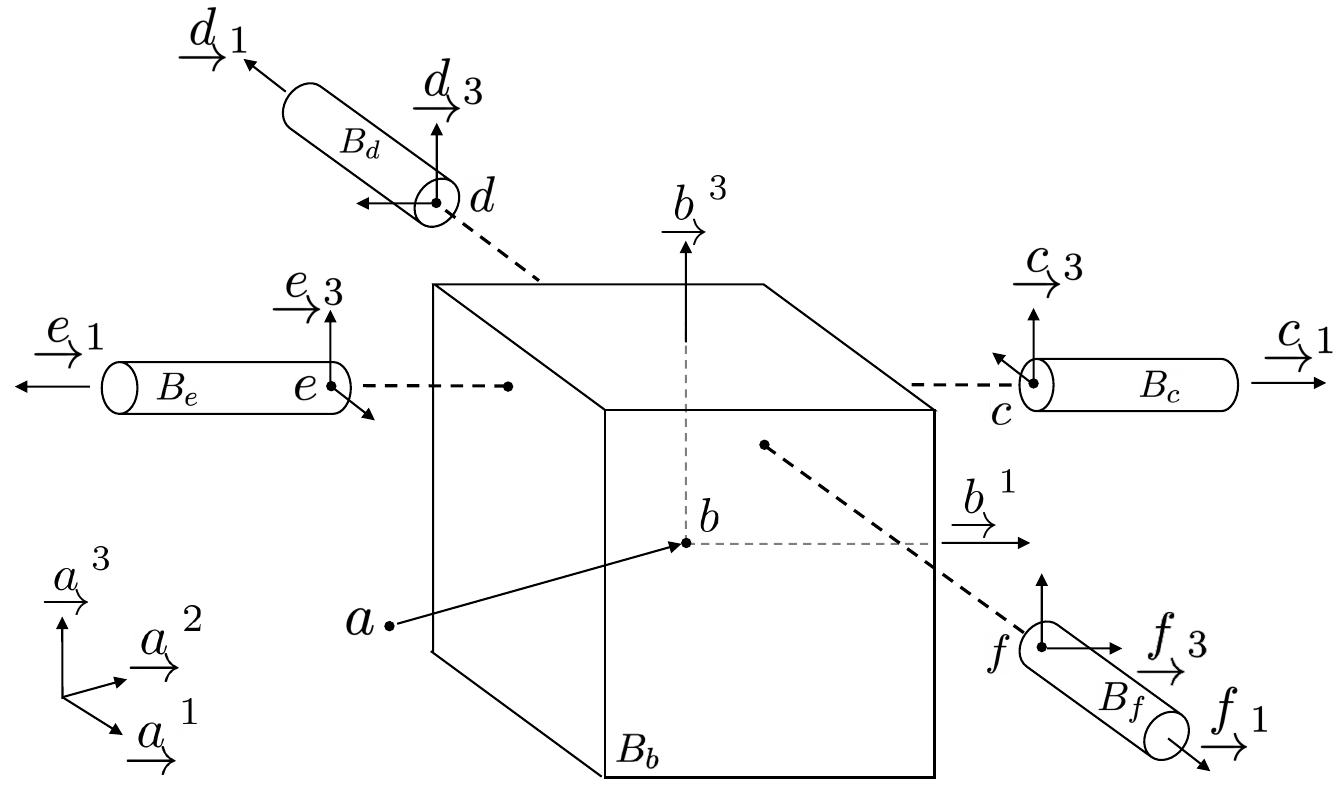}
    \caption{ Exploded view of CABLESSail Assembly~\#1 consisting of solar sail bus $B_b$, and CABLESSail booms $B_c$ - $B_f$, shortened for clarity, with corresponding body frames and attachment points.}
    \label{fig:Stubby Sail Assembly}
\end{figure}

\subsection{Aggregate Component Equations of Motion}
The unconstrained dynamics for each individual component are gathered into a single set of equations to fully represent the system of unassembled components. The generalized coordinates of this aggregate system are $\mbf{q}^\trans = 
    \Big[
        \mbf{q}^{B_b^\trans} \,\,\,
        \mbf{q}^{B_c^\trans} \,\,\,
        \mbf{q}^{B_d^\trans} \,\,\,
        \mbf{q}^{B_e^\trans} \,\,\,
        \mbf{q}^{B_f^\trans}
    \Big]$,
while the generalized velocities of the aggregate system are $\mbs{\nu}^\trans = 
    \Big[
        \mbs{\nu}^{B_b^\trans} \,\,\,
        \mbs{\nu}^{B_c^\trans} \,\,\,
        \mbs{\nu}^{B_d^\trans} \,\,\,
        \mbs{\nu}^{B_e^\trans} \,\,\,
        \mbs{\nu}^{B_f^\trans}
    \Big]$.  The unconstrained dynamics of the components are written in an aggregate form as
\begin{equation}\label{eq: full unconstrained system dynamics}
    \mbf{M} \dot{\mbs{\nu}} +
    \mbs{\Gamma}^{\trans} \mbf{K} \mbf{q}
    + \mbf{f}^{non}
    =
    \mbs{\Gamma}^{\trans}\mbf{f},
\end{equation}
where the aggregate matrices are $\mbf{M} = \text{diag} 
    \left\{  \mbf{M}_{B_b} , \mbf{M}_{B_c} , \mbf{M}_{B_d} , \mbf{M}_{B_e} , \mbf{M}_{B_f}   \right\}$, $
    \mbs{\Gamma} = \text{diag}
    \left\{ \mbs{\Gamma}_{B_b}, \mbs{\Gamma}_{B_c}, \mbs{\Gamma}_{B_d}, \mbs{\Gamma}_{B_e} , \mbs{\Gamma}_{B_f} \right\}$, 
    $\mbf{K} = \text{diag}
    \left\{ \mbf{K}_{B_b}, \mbf{K}_{B_c}, \mbf{K}_{B_d}, \mbf{K}_{B_e}, \mbf{K}_{B_f} \right\}$, 
    $\mbs{\Omega} = \text{diag}
    \left\{  \mbs{\Omega}_{B_b},  \mbs{\Omega}_{B_c},  \mbs{\Omega}_{B_d},  \mbs{\Omega}_{B_e},  \mbs{\Omega}_{B_f} \right\}$, 
    $\mbf{f}^\trans = 
    \Big[
        \mbf{f}_{B_b}^\trans \,\,\, \mbf{f}_{B_c}^\trans \,\,\, \mbf{f}_{B_d}^\trans \,\,\, \mbf{f}_{B_e}^\trans \,\,\, \mbf{f}_{B_f}^\trans
    \Big]$,
    and 
    $\mbf{f}^{non^\trans} = 
    \Big[
        \mbf{f}^{non^\trans}_{B_b} \,\,\, \mbf{f}^{non^\trans}_{B_c} \,\,\, \mbf{f}^{non^\trans}_{B_d} \,\,\, \mbf{f}^{non^\trans}_{B_e} \,\,\, \mbf{f}^{non^\trans}_{B_f}
    \Big]$.

\subsection{Component Assembly Constraints}
The base of each boom, $B_i, \; i \in \left\{ c, d, e, f\right\}$, is constrained to the perimeter of the spacecraft hub, $B_b$. This is expressed through the holonomic constraint
\begin{equation}\label{Position_Holonomic_Constraint}
    \mbf{r}^{ia}_a = \mbf{r}^{ba}_a + \mbf{C}_{ab}\mbf{r}^{ib}_b
    ,
\end{equation}
where $\mbf{r}^{ib}_b$ describes the constant position of the boom attachment point on the spacecraft hub relative to its center of mass. Differentiating \eqref{Position_Holonomic_Constraint} with respect to time results in
\begin{equation*}
    \dot{\mbf{r}}_a^{ia} = \dot{\mbf{r}}_a^{ba} - \mbf{C}_{ab}\mbf{r}_b^{ib^\times}\mbs{\omega}^{ba}_b,
\end{equation*}
which is used to express the constraint in Pfaffian form as
\begin{equation}\label{eq:Pfaffian form constraint velocities}
    \dot{\mbf{r}}_a^{ia} - \dot{\mbf{r}}_a^{ba} + \mbf{C}_{ab}\mbf{r}_b^{ib^\times}\mbs{\omega}^{ba}_b = \mbf{0}.
\end{equation}

To constrain the booms to be rigidly attached to the spacecraft hub, i.e., the boom base cannot rotate with respect to the hub, the angular velocities of each body relative to an inertial reference frame must be equal ($\mbs{\omega}^{ba}_i = \mbs{\omega}^{ia}_i$). This leads to the following constraint in Pfaffian form,
\begin{equation}\label{eq:Pfaffian form constraint omeags}
\mbs{\omega}^{ia}_i - \mbs{\omega}^{ba}_i = \mbf{0}.
\end{equation}

All of the assembly constraints are then expressed as $\mbs{\Xi} \mbs{\nu} = \mbf{0}$, where
\begin{equation} \label{Xi_full_constraint_matrix}
\mbs{\Xi}=
    \begin{bmatrix}
        -\mbf{1} & \mbf{C}_{ab}\mbf{r}_b^{cb^\times} & \mbf{0} & \mbf{1} & \mbf{0} & \mbf{0} & \mbf{0} & \mbf{0} & \mbf{0} & \mbf{0} & \mbf{0} & \mbf{0} & \mbf{0} & \mbf{0} \\
        -\mbf{1} & \mbf{C}_{ab}\mbf{r}_b^{db^\times} & \mbf{0} & \mbf{0} & \mbf{0} & \mbf{0} & \mbf{1} & \mbf{0} & \mbf{0} & \mbf{0} & \mbf{0} & \mbf{0} & \mbf{0} & \mbf{0} \\
        -\mbf{1} & \mbf{C}_{ab}\mbf{r}_b^{eb^\times} & \mbf{0} & \mbf{0} & \mbf{0} & \mbf{0} & \mbf{0} & \mbf{0} & \mbf{0} & \mbf{1} & \mbf{0} & \mbf{0} & \mbf{0} & \mbf{0} \\
        -\mbf{1} & \mbf{C}_{ab}\mbf{r}_b^{fb^\times} & \mbf{0} & \mbf{0} & \mbf{0} & \mbf{0} & \mbf{0} & \mbf{0} & \mbf{0} & \mbf{0} & \mbf{0} & \mbf{0} & \mbf{1} & \mbf{0} \\ 
        \mbf{0}  & - \mbf{C}_{cb} & \mbf{0} & \mbf{1} & \mbf{0} & \mbf{0} & \mbf{0} & \mbf{0} & \mbf{0} & \mbf{0} & \mbf{0} & \mbf{0} & \mbf{0} & \mbf{0} \\
        \mbf{0}  & - \mbf{C}_{db} & \mbf{0} & \mbf{0} & \mbf{0} & \mbf{0} & \mbf{1} & \mbf{0} & \mbf{0} & \mbf{0} & \mbf{0} & \mbf{0} & \mbf{0} & \mbf{0} \\
        \mbf{0}  & - \mbf{C}_{eb} & \mbf{0} & \mbf{0} & \mbf{0} & \mbf{0} & \mbf{0} & \mbf{0} & \mbf{0} & \mbf{1} & \mbf{0} & \mbf{0} & \mbf{0} & \mbf{0} \\ 
        \mbf{0}  & - \mbf{C}_{fb} & \mbf{0} & \mbf{0} & \mbf{0} & \mbf{0} & \mbf{0} & \mbf{0} & \mbf{0} & \mbf{0} & \mbf{0} & \mbf{0} & \mbf{1} & \mbf{0} \\ 
    \end{bmatrix}.
\end{equation}
The assembly constraints are added to the aggregate equations of motion in~\eqref{eq: full unconstrained system dynamics} to obtain the system's constrained equations of motion 
\begin{equation}\label{eq: full unconstrained system dynamics2}
    \mbf{M} \dot{\mbs{\nu}} +
    \mbs{\Gamma}^{\trans} \mbf{K} \mbf{q}
    + \mbf{f}^{non}
    =
    \mbs{\Gamma}^{\trans}\mbf{f} + \mbs{\Xi}^\trans \mbs{\lambda}
    ,
\end{equation}
where $\mbs{\lambda}$ contains the Lagrange multipliers that maintain the assembly constraint. Interestingly, the use of the null-space method in the following section will remove the need to compute both the Lagrange multipliers $\mbs{\lambda}$ and the constraint matrix $\mbs{\Xi}$ when numerically solving these equations of motion in simulation.

\subsection{Change of Variables}
A reduced set of augmented velocities is chosen as 
$\hat{\mbs{\nu}}^\trans = 
    \Big[
        \dot{\mbf{r}}^{ba^\trans}_a \,\,\,
        \mbs{\omega}^{ba^\trans}_b \,\,\,
        \dot{\mbf{q}}^{\epsilon c^\trans} \,\,\,
        \dot{\mbf{q}}^{\epsilon d^\trans} \,\,\,
        \dot{\mbf{q}}^{\epsilon e^\trans} \,\,\,
        \dot{\mbf{q}}^{\epsilon f^\trans}
    \Big]$,
which represent a minimal set of generalized velocities needed to represent the constrained motion of the assembled system. 
The augmented velocities are related to the unconstrained generalized velocities through the Pfaffian form constraints,~\eqref{eq:Pfaffian form constraint velocities} and~\eqref{eq:Pfaffian form constraint omeags}, and summarized by the relationship $\mbs{\nu} = \mbs{\Upsilon}\hat{\mbs{\nu}}$. The matrix $\mbs{\Upsilon}$ and its derivative with respect to time are given by
\begin{equation}
\label{eq:Upsilon}
       \mbs{\Upsilon}^\trans =  \begin{bmatrix}
        \mbf{1} & \mbf{0} & \mbf{1} & \mbf{0} & \mbf{0} & \mbf{1} & \mbf{0} & \mbf{0} & \mbf{1} & \mbf{0} & \mbf{0} & \mbf{1} & \mbf{0} & \mbf{0}  \\
        \mbf{0} & \mbf{1} & \mbf{r}_b^{cb^\times}\mbf{C}_{ba} & \mbf{C}_{bc} & \mbf{0} & 
        \mbf{r}_b^{db^\times}\mbf{C}_{ba} & \mbf{C}_{bd} & \mbf{0} & 
        \mbf{r}_b^{eb^\times}\mbf{C}_{ba} & \mbf{C}_{be} & \mbf{0} & 
        \mbf{r}_b^{fb^\times}\mbf{C}_{ba} & \mbf{C}_{bf} & \mbf{0} \\
        \mbf{0} & \mbf{0} & \mbf{0} & \mbf{0} & \mbf{1} & \mbf{0} & \mbf{0} & \mbf{0} & \mbf{0} & \mbf{0} & \mbf{0} & \mbf{0} & \mbf{0} & \mbf{0} \\
        \mbf{0} & \mbf{0} & \mbf{0} & \mbf{0} & \mbf{0} & \mbf{0} & \mbf{0} & \mbf{1} & \mbf{0} & \mbf{0} & \mbf{0} & \mbf{0} & \mbf{0} & \mbf{0} \\
        \mbf{0} & \mbf{0} & \mbf{0} & \mbf{0} & \mbf{0} & \mbf{0} & \mbf{0} & \mbf{0} & \mbf{0} & \mbf{0} & \mbf{1} & \mbf{0} & \mbf{0} & \mbf{0} \\
        \mbf{0} & \mbf{0} & \mbf{0} & \mbf{0} & \mbf{0} & \mbf{0} & \mbf{0} & \mbf{0} & \mbf{0} & \mbf{0} & \mbf{0} & \mbf{0} & \mbf{0} & \mbf{1}
        \end{bmatrix},        
\end{equation}
\begin{equation}
\label{eq:UpsilonDot}
       \dot{\mbs{\Upsilon}}^\trans =  \begin{bmatrix}
        \mbf{0} & \mbf{0} & \mbf{0} & \mbf{0} & \mbf{0} & \mbf{0} & \mbf{0} & \mbf{0} & \mbf{0} & \mbf{0} & \mbf{0} & \mbf{0} & \mbf{0} & \mbf{0}  \\
        \mbf{0} & \mbf{0} & -\mbf{r}_b^{cb^\times}\mbs{\omega}_b^{ba^\times}\mbf{C}_{ba} & \mbf{0} & \mbf{0} & 
        -\mbf{r}_b^{db^\times}\mbs{\omega}_b^{ba^\times}\mbf{C}_{ba} & \mbf{0} & \mbf{0} & 
        -\mbf{r}_b^{eb^\times}\mbs{\omega}_b^{ba^\times}\mbf{C}_{ba} & \mbf{0} & \mbf{0} & 
        -\mbf{r}_b^{fb^\times}\mbs{\omega}_b^{ba^\times}\mbf{C}_{ba} & \mbf{0} & \mbf{0} \\
        \mbf{0} & \mbf{0} & \mbf{0} & \mbf{0} & \mbf{0} & \mbf{0} & \mbf{0} & \mbf{0} & \mbf{0} & \mbf{0} & \mbf{0} & \mbf{0} & \mbf{0} & \mbf{0}  \\
        \mbf{0} & \mbf{0} & \mbf{0} & \mbf{0} & \mbf{0} & \mbf{0} & \mbf{0} & \mbf{0} & \mbf{0} & \mbf{0} & \mbf{0} & \mbf{0} & \mbf{0} & \mbf{0} \\
        \mbf{0} & \mbf{0} & \mbf{0} & \mbf{0} & \mbf{0} & \mbf{0} & \mbf{0} & \mbf{0} & \mbf{0} & \mbf{0} & \mbf{0} & \mbf{0} & \mbf{0} & \mbf{0} \\
        \mbf{0} & \mbf{0} & \mbf{0} & \mbf{0} & \mbf{0} & \mbf{0} & \mbf{0} & \mbf{0} & \mbf{0} & \mbf{0} & \mbf{0} & \mbf{0} & \mbf{0} & \mbf{0} 
        \end{bmatrix}.        
\end{equation}
It can be shown that $\mbs{\Xi} \mbs{\Upsilon} = \mbf{0}$ by substituting $\mbs{\nu} = \mbs{\Upsilon}\hat{\mbs{\nu}}$ into $\mbs{\Xi} \mbs{\nu} = \mbf{0}$, which is the central property of the null-space method.

\subsection{Constrained Solar Sail Equations of Motion: The Null-Space Method} \label{sec:Constrained Solar Sail Equations of Motion}
The null-space method is applied to the equations of motion in~\eqref{eq: full unconstrained system dynamics2} by first pre-multipllying on the left by $\mbs{\Upsilon}^\trans$, resulting in
\begin{equation}
\label{eq:EoMNullSpaceTotal1}
    \mbs{\Upsilon}^\trans\mbf{M} \dot{\mbs{\nu}} +
    \mbs{\Upsilon}^\trans\mbs{\Gamma}^{\trans} \mbf{K} \mbf{q}
    + \mbs{\Upsilon}^\trans\mbf{f}^{non}
    =
    \mbs{\Upsilon}^\trans\mbs{\Gamma}^{\trans}\mbf{f}
    + \mbs{\Upsilon}^\trans \mbs{\Xi}^\trans \mbs{\lambda}
    .
\end{equation}
With $\mbs{\Upsilon}^\trans \mbs{\Xi}^\trans = \mbf{0}$, the Lagrange multipliers are now removed from~\eqref{eq:EoMNullSpaceTotal1}.
The time derivative of the change of coordinate relationship, $\dot{\mbs{\nu}} = \mbs{\Upsilon} \dot{\mbshat{\nu}} + \dot{\mbs{\Upsilon}} \mbshat{\nu}$, is then substituted into~\eqref{eq:EoMNullSpaceTotal1}, resulting in
\begin{equation*}
    \mbs{\Upsilon}^\trans\mbf{M} (\mbs{\Upsilon}\dot{\hat{\mbs{\nu}}} + \dot{\mbs{\Upsilon}} \hat{\mbs{\nu}}) +
    \mbs{\Upsilon}^\trans\mbs{\Gamma}^{\trans} \mbf{K} \mbf{q}
    + \mbs{\Upsilon}^\trans\mbf{f}^{non}
    =
    \mbs{\Upsilon}^\trans\mbs{\Gamma}^{\trans}\mbf{f}
    ,
\end{equation*}
which can be written in the compact form
\begin{equation} \label{eq: final equaiton of motion after null space}
    \bar{\mbf{M}} \dot{\hat{\mbs{\nu}}} + \Bar{\mbf{K}} \mbf{q} + \Bar{ \mbf{f} }_{non} = \bar{\mbf{f}},
\end{equation}
where $\bar{\mbf{M}} = \mbs{\Upsilon}^\trans\mbf{M} \mbs{\Upsilon}$, $\Bar{\mbf{K}} = \mbs{\Upsilon}^\trans\mbs{\Gamma}^{\trans} \mbf{K}$, $\bar{\mbf{f}}_{non} = \mbs{\Upsilon}^\trans\mbf{f}^{non} + \mbs{\Upsilon}^\trans\mbf{M} \dot{\mbs{\Upsilon}} \hat{\mbs{\nu}}$, and $\bar{\mbf{f}} = \mbs{\Upsilon}^\trans\mbs{\Gamma}^{\trans}\mbf{f}$.

\par It is worth emphasizing the modularity of equations of motion derived in~\eqref{eq: final equaiton of motion after null space}. For example, if a component is to be altered in the assembly or the constrained is to be adjusted, the only changes required are to the constraint matrices, $\mbs{\Upsilon}$ and $\dot{\mbs{\Upsilon}}$, and updating the component sub-matrices within the aggregate matrices $\mbf{M}$, $\mbf{K}$, $\mbs{\Gamma}$,etc. with those from the new component. This modularity is reflected in the structure of the numerical simulation code for this system.

\subsection{Second Assembly with Active Mass Translator}
\label{sec:Second Assembly with Active Mass Translator}

\par A second solar sail component configuration is used in this work and is a modification of the previous model achieved with the null-space method. The previous configuration is modified by including an additional component, the spacecraft bus $\mathcal{B}_g$, and modifying an existing component, the solar sail bus $\mathcal{B}_b$. This is done to model the effect of an active mass translator actuator~\cite{inness2023momentum}, which has the ability to translate two spacecraft bus components relative to one another. To include the new component the corresponding matrices, generalized coordinates, generalized velocities, and vectors describing the component dynamics from Section~\ref{sec: spacecraft bus energies} are simply appended to the existing aggregated assembly dynamics $\mbf{M}, \mbf{K}, \mbs{\Gamma}, \mbs{\Omega}, \mbf{f}, $ and $\mbf{f}^{non}$. Two additional rows are inserted into $\mbs{\Xi}$ to enforce the new constraint assembling the new spacecraft bus onto the existing solar sail bus, which are 
\begin{align*}
     & \dot{\mbf{r}}_a^{ga} - \dot{\mbf{r}}_a^{ba} + \mbf{C}_{ab}\mbf{r}_b^{gb^\times}\mbs{\omega}^{ba}_b = \mbf{0},
\\
     & \mbs{\omega}^{ga}_g - \mbs{\omega}^{ba}_g = \mbf{0} .
\end{align*}


\begin{figure}[t!]
    \centering
    \includegraphics[scale=0.35]{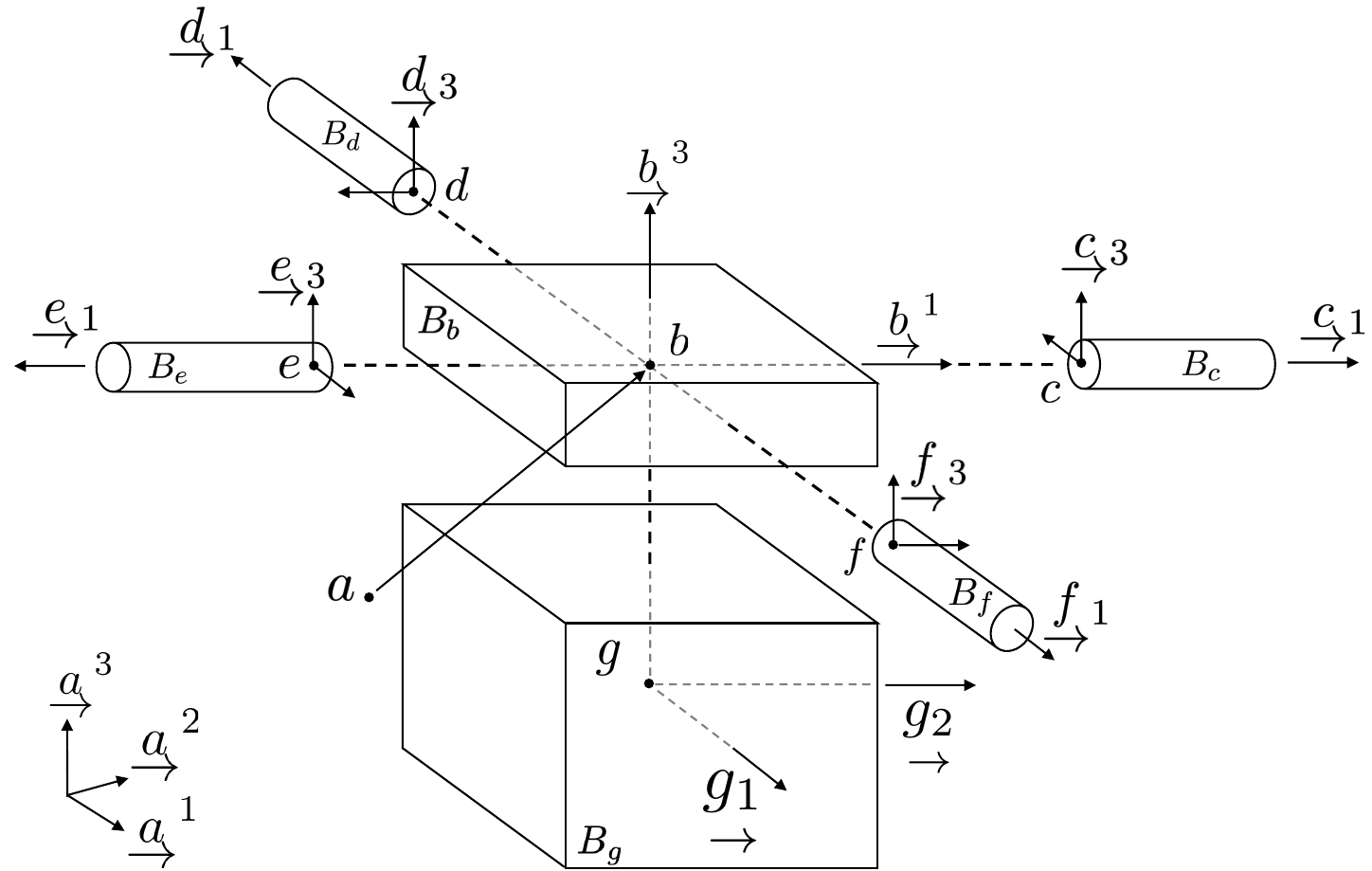}
    \caption{ Exploded view of CABLESSail Assembly~\#2 consisting of solar sail bus $B_b$, spacecraft bus $B_g$, and CABLESSail booms $B_c$ - $B_f$, shortened for clarity, with corresponding body frames and attachment points.}
    \label{fig:Assembly two exploded view}
\end{figure}



The constraint matrix, $\mbs \Xi$, is modified from Assembly \#1 with additional zero columns appended to account for the increased length of $\mbs{\nu}$, and the additions of row five and ten that rigidly constrain the spacecraft bus, $B_g$, to the solar sail bus $B_b$ in translation and rotation, respectively. The actuation position of the AMT is encapsulated within $\mbf{r}_b^{gb}$, which describes the position of the spacecraft bus center of mass relative to the solar sail bus center of mass. The form of $\mbs \Xi$ for this case is given as
\begin{equation*} \label{Xi_full_constraint_matrix2}
\mbs{\Xi}=
    \begin{bmatrix}
        -\mbf{1} & \mbf{C}_{ab}\mbf{r}_b^{cb^\times} & \mbf{0} & \mbf{1} & \mbf{0} & \mbf{0} & \mbf{0} & \mbf{0} & \mbf{0} & \mbf{0} & \mbf{0} & \mbf{0} & \mbf{0} & \mbf{0} & \mbf{0} & \mbf{0} \\
        -\mbf{1} & \mbf{C}_{ab}\mbf{r}_b^{db^\times} & \mbf{0} & \mbf{0} & \mbf{0} & \mbf{0} & \mbf{1} & \mbf{0} & \mbf{0} & \mbf{0} & \mbf{0} & \mbf{0} & \mbf{0} & \mbf{0} & \mbf{0} & \mbf{0} \\
        -\mbf{1} & \mbf{C}_{ab}\mbf{r}_b^{eb^\times} & \mbf{0} & \mbf{0} & \mbf{0} & \mbf{0} & \mbf{0} & \mbf{0} & \mbf{0} & \mbf{1} & \mbf{0} & \mbf{0} & \mbf{0} & \mbf{0} & \mbf{0} & \mbf{0} \\
        -\mbf{1} & \mbf{C}_{ab}\mbf{r}_b^{fb^\times} & \mbf{0} & \mbf{0} & \mbf{0} & \mbf{0} & \mbf{0} & \mbf{0} & \mbf{0} & \mbf{0} & \mbf{0} & \mbf{0} & \mbf{1} & \mbf{0} & \mbf{0} & \mbf{0} \\ 
        -\mbf{1} & \mbf{C}_{ab}\mbf{r}_b^{gb^\times} & \mbf{0} & \mbf{0} & \mbf{0} & \mbf{0} & \mbf{0} & \mbf{0} & \mbf{0} & \mbf{0} & \mbf{0} & \mbf{0} & \mbf{1} & \mbf{0} & \mbf{0} & \mbf{0} \\
        \mbf{0}  & - \mbf{C}_{cb} & \mbf{0} & \mbf{1} & \mbf{0} & \mbf{0} & \mbf{0} & \mbf{0} & \mbf{0} & \mbf{0} & \mbf{0} & \mbf{0} & \mbf{0} & \mbf{0} & \mbf{0} & \mbf{0} \\
        \mbf{0}  & - \mbf{C}_{db} & \mbf{0} & \mbf{0} & \mbf{0} & \mbf{0} & \mbf{1} & \mbf{0} & \mbf{0} & \mbf{0} & \mbf{0} & \mbf{0} & \mbf{0} & \mbf{0} & \mbf{0} & \mbf{0} \\
        \mbf{0}  & - \mbf{C}_{eb} & \mbf{0} & \mbf{0} & \mbf{0} & \mbf{0} & \mbf{0} & \mbf{0} & \mbf{0} & \mbf{1} & \mbf{0} & \mbf{0} & \mbf{0} & \mbf{0} & \mbf{0} & \mbf{0} \\ 
        \mbf{0}  & - \mbf{C}_{fb} & \mbf{0} & \mbf{0} & \mbf{0} & \mbf{0} & \mbf{0} & \mbf{0} & \mbf{0} & \mbf{0} & \mbf{0} & \mbf{0} & \mbf{1} & \mbf{0} & \mbf{0} & \mbf{0} \\
        \mbf{0}  & - \mbf{C}_{gb} & \mbf{0} & \mbf{0} & \mbf{0} & \mbf{0} & \mbf{0} & \mbf{0} & \mbf{0} & \mbf{0} & \mbf{0} & \mbf{0} & \mbf{1} & \mbf{0} & \mbf{0} & \mbf{0} \\
    \end{bmatrix}.
\end{equation*}

\par The change of variables is modified from Assembly \#2 by appending two additional rows to the matrices $\mbs{\Upsilon}$ and $\dot{\mbs{\Upsilon}}$ defined in~\eqref{eq:Upsilon} and~\eqref{eq:UpsilonDot}. Specifically, the additional rows appended to the end of $\mbs{\Upsilon}$ are given by
    \begin{equation*}
    \begin{bmatrix}
        \mbf{1} & -\mbf{C}_{ab}\mbf{r}_b^{gb^\times}   & \mbf{0} & \mbf{0} & \mbf{0} & \mbf{0} \\
        \mbf{0} & \mbf{C}_{gb}                         & \mbf{0} & \mbf{0} & \mbf{0} & \mbf{0} 
    \end{bmatrix},
\end{equation*}
while the rows appended to $\dot{\mbs{\Upsilon}}$ are
\begin{equation*}
    \begin{bmatrix}
                \mbf{0} & -\mbf{C}_{ab}\mbs{\omega}_b^{ba^\times}\mbf{r}_b^{gb^\times} & \mbf{0} & \mbf{0} & \mbf{0} & \mbf{0} \\
        \mbf{0} & \mbf{0}                                                      & \mbf{0} & \mbf{0} & \mbf{0} & \mbf{0} \\
    \end{bmatrix}.
\end{equation*}

\section*{Funding Sources}
This work was supported by an Early Career Faculty grant from NASA’s Space Technology Research Grants Program under award No.~80NSSC23K0075.

\bibliography{Bib}

\end{document}